\documentclass[acs,preprint,showkeys]{revtex4-1}
\usepackage{graphicx}

\begin{document}

\title{Structure and Dynamics of [PF$_6$][P$_{1,2,2,4}$] from Molecular Dynamics Simulations}

\author{Marcelo A. Carignano}
\email{mcarignano@qf.org.qa}
\affiliation{Qatar Environment and Energy Research Institute, P.O. Box 5825, Doha, Qatar}

\date{November 19, 2013}

\begin{abstract}
Diethyl(methyl)(isobutyl)phosphonium hexafluorophosphate, [PF$_6$][P$_{1,2,2,4}$], is an organic ionic plastic crystal with potential uses as a solid electrolyte in storage and light harvesting devices. In this work we present a molecular dynamics simulation study for this material covering an extended temperature range, from 175 K to 500 K. The simulations predicts a transition from the crystalline to a {\em semi} plastic phase at 197 K, the onset of cation jump-like rotations at 280 K, a third transition at 340 K to a {\em full} plastic phase and melting to 450 K. Overall, the simulations show a good agreement with the experimental findings providing a wealth of detail in the structural and dynamic properties of the system.
\end{abstract}


\maketitle

\newpage

\section{Introduction}

Research addressing solid electrolytes and ionic conductors has surged in the recent years due to the ever increasing need for storage devices with higher energy density \cite{Armand:2008aa,Braun:2012aa}. Organic ionic plastic crystals (OIPC) are an attractive alternative for the electrolyte of lithium ion batteries since they offer good thermal stability, low vapor pressure, non-flammability, and good mechanical properties that allow their shaping even in thin films \cite{Pringle:2010aa,Pringle:2013aa}. At an experimental level, there have been several applications of OIPC in lithium ion batteries \cite{Howlett:2011aa,MacFarlane:1999aa,Alarco:2004aa} and dye sensitized solar cells \cite{Li:2012aa}. The distinguishable characteristic that defines plastic crystals is that they posses long range translation order while exhibiting a local rotational disorder \cite{Sherwood:1979aa}. The simplest plastic crystals are perhaps halomethane compounds, which have no use for solid-state electronics that we are aware of, but are interesting in order to develop a computational methodology to study plastic crystal in general \cite{Zuriaga:2011aa,Caballero:2012aa,Caballero:2013aa} and understand their behavior at the atomic level. For example, CCl$_3$Br exhibit a crystalline phase at low temperature, transforms to a plastic phase at 160 K and melts at 350 K. OIPC offer a richer phase behavior, mainly due to the fact that they are composed by an anion-cation dimer that are often of different size and molecular architecture. This asymmetry allows the emergence of a {\em semi} plastic phase as one of the monomers is able to rotate, and eventually a second {\em full} plastic phase as the second component gain rotational degrees of freedom. An example of an OIPC recently characterized by experimental techniques is diethyl(methyl)(isobutyl)phosphonium hexafluorophosphate, or [PF$_6$][P$_{1,2,2,4}$] in short. Jin et al. \cite{Jin:2012aa} recently published an exhaustive report on the structure, thermodynamic and transport properties of [PF$_6$][P$_{1,2,2,4}$]. The picture presented in Ref \cite{Jin:2012aa} consists of five distinct phases from crystalline to melt: the {\em semi} plastic phase IV, characterized by rotations of the smaller component PF$_6$, transforms at 298 K to phase III in which directional rotations of the larger component are visible. At 343 K full rotations of P$_{1,2,2,4}$ are observed along with diffusion of the PF$_6$. Transformation to phase I occurs at 393 K in which diffusion of both ions is appreciable. The system melts at 423 K. 

In this paper we present an extensive simulation study of [PF$_6$][P$_{1,2,2,4}$] in a broad temperature range, trying to capture the different phases observed experimentally and characterize them in terms of the atomistic structure and dynamics. For that purpose, we use a force field especially developed for ionic liquids \cite{Lopes:2004aa,Lopes:2004ab,Canongia-Lopes:2006aa,Lopes:2008aa,Canongia-Lopes:2012aa}, which is an evolution of the OPLS-AA force field \cite{Jorgensen:1996aa}. The phase transformations reported in Ref. \cite{Jin:2012aa} occur in a range of temperature spanning over 300 K. To capture the behavior of the system over such a broad temperature range using atomistic simulations based on one single model represents a challenge and a strong test for a force field that has been developed to reproduce experimental results at standard thermodynamic conditions. Yet, in order to properly understand the energy scale associated with a model it is very important to perform tests over a broad set of conditions. This step is essential in order to improve our overall understanding in order to guide the development of better atomistic descriptions of real systems. In this work we found that the CL\&P force field, used with an overall charge rescaling factor of 0.8, perform remarkably well and provides a very good description of the system in line with the known experimental data.

Upon the completion of this work it was brought to our attention a new simulation study \cite{Chen:2013aa} on the same system recently published by the same group that performed the experimental study. Therefore, our work overlaps significantly with Ref. \cite{Chen:2013aa}. Nevertheless, the two works are based on different force fields, and our simulations that are on a larger system and covering much longer simulation times provide a more detailed description of the different molecular processes involved in the transitions between the different regimes. This paper is organized as follows: Section \ref{model} describe the model system and details of the simulation methodology. Section \ref{results} presents and discuss the results from a detailed molecular perspective. An extensive Supplemental Information file is included to further clarify our methodology and findings. Finally, in Section \ref{final} we present a general summary and discussion.

\section{Models and Methods}
\label{model}

\begin{figure}[!t]
\includegraphics*[width=0.28\textwidth]{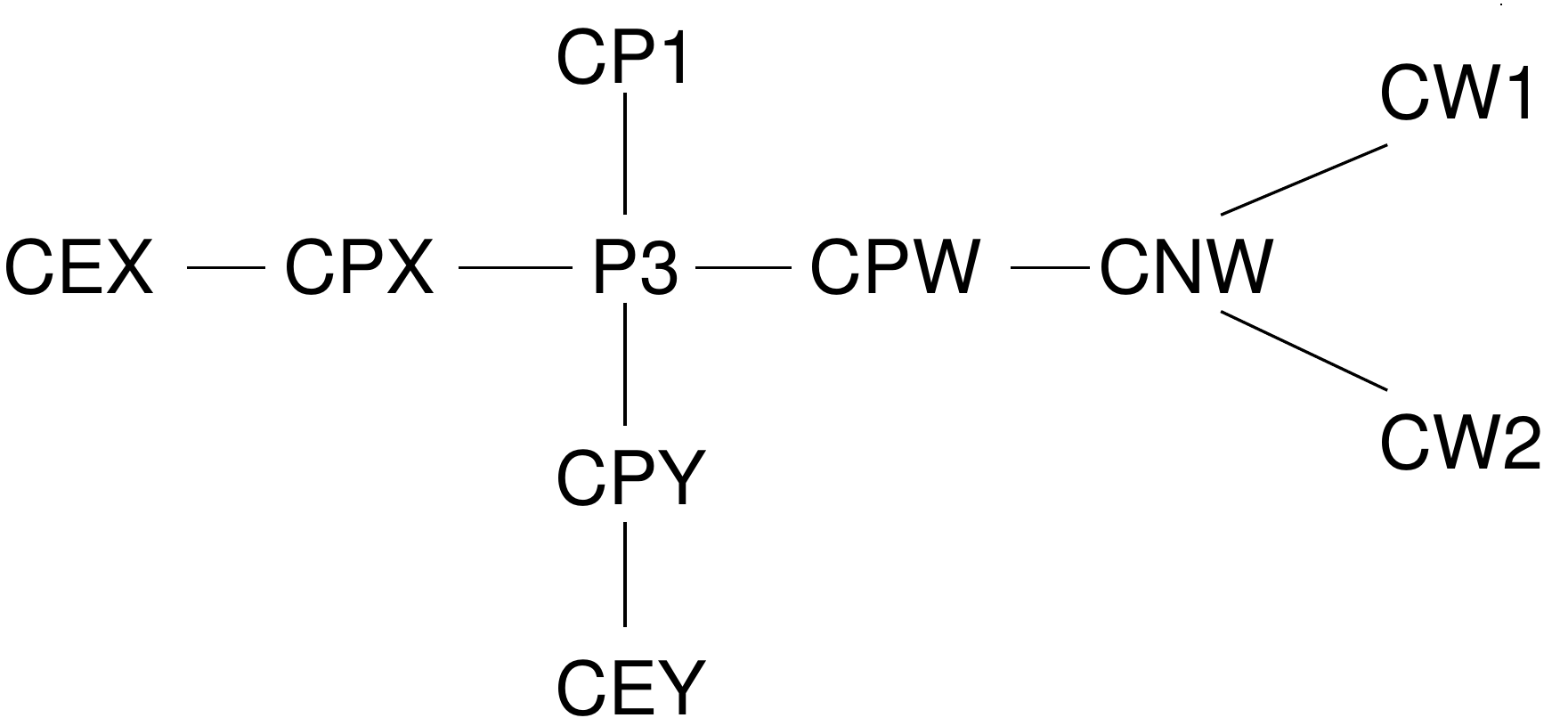}
\caption{Labeling scheme used for the [P$_{1,2,2,4}$]. The CH$_3$ groups are referred to after their C atom as M1, MX, MY, MW1 and MW2. The atoms of the anion are labeled with P and F$_i$, with $i=1,\cdots,6$.}
\label{labeler}
\end{figure}

The OIPC system that we target in this study is diethyl(methyl)(isobutyl)phosphonium hexafluorophosphate, [PF$_6$][P$_{1,2,2,4}$], which has the chemical structure displayed in the top left panel of Figure \ref{views}. In order to facilitate the description of the results presented below, it is convenient to introduce a labeling scheme to refer to the different  atoms univocally. Our labeling choice is depicted in Figure \ref{labeler}.

\begin{figure}[!t]
\includegraphics*[width=0.22\textwidth]{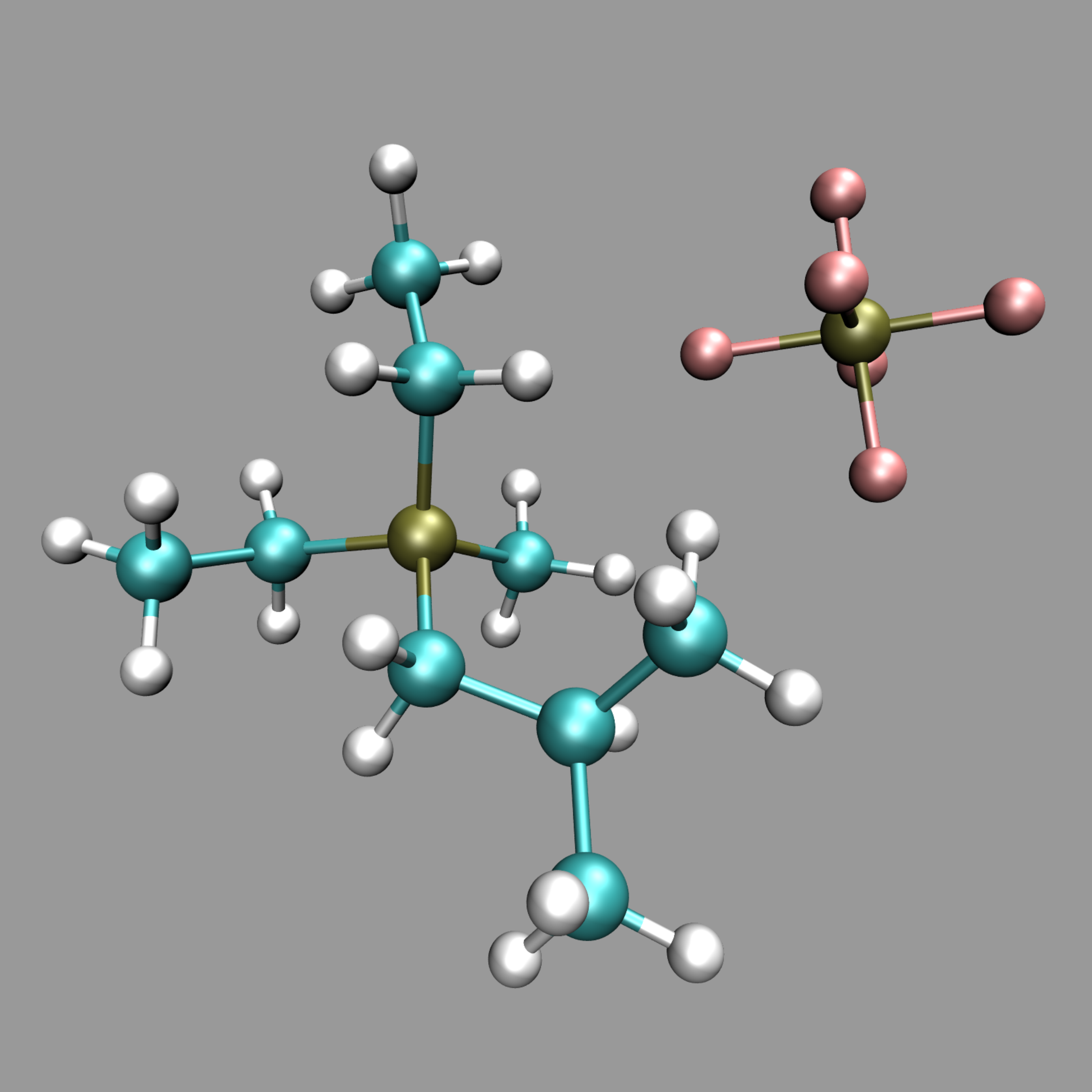}
\includegraphics*[width=0.22\textwidth]{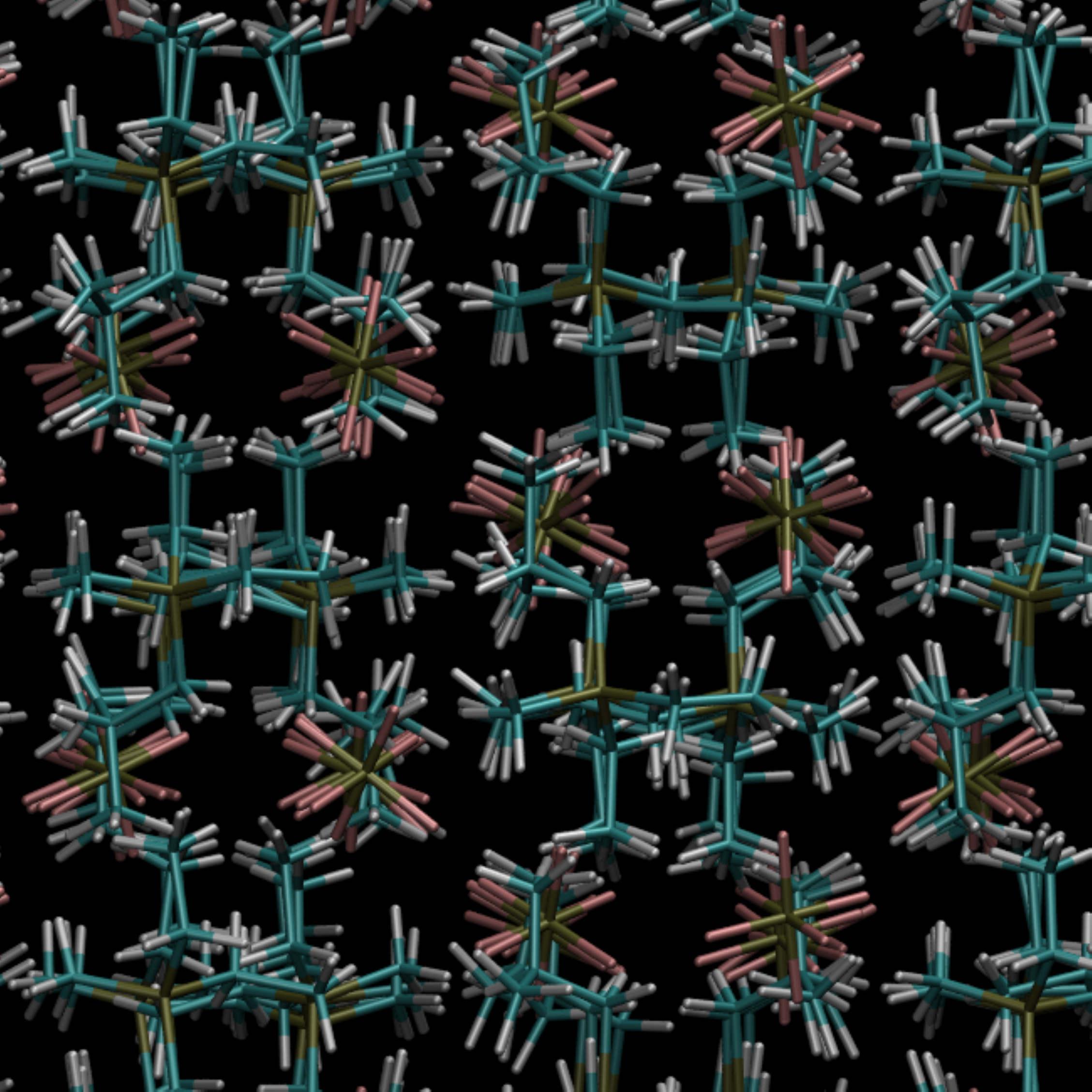}\\ 
\includegraphics*[width=0.22\textwidth]{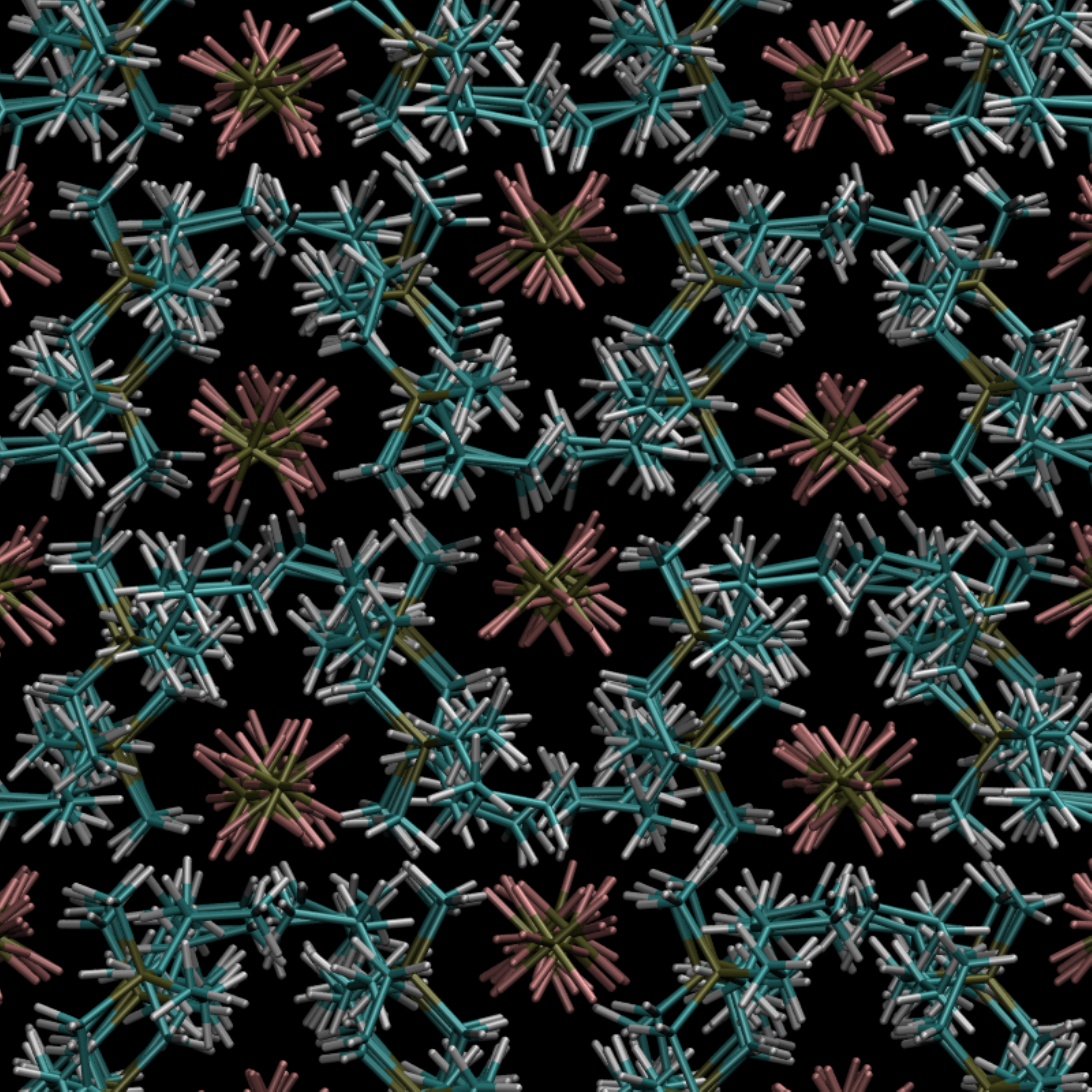}
\includegraphics*[width=0.22\textwidth]{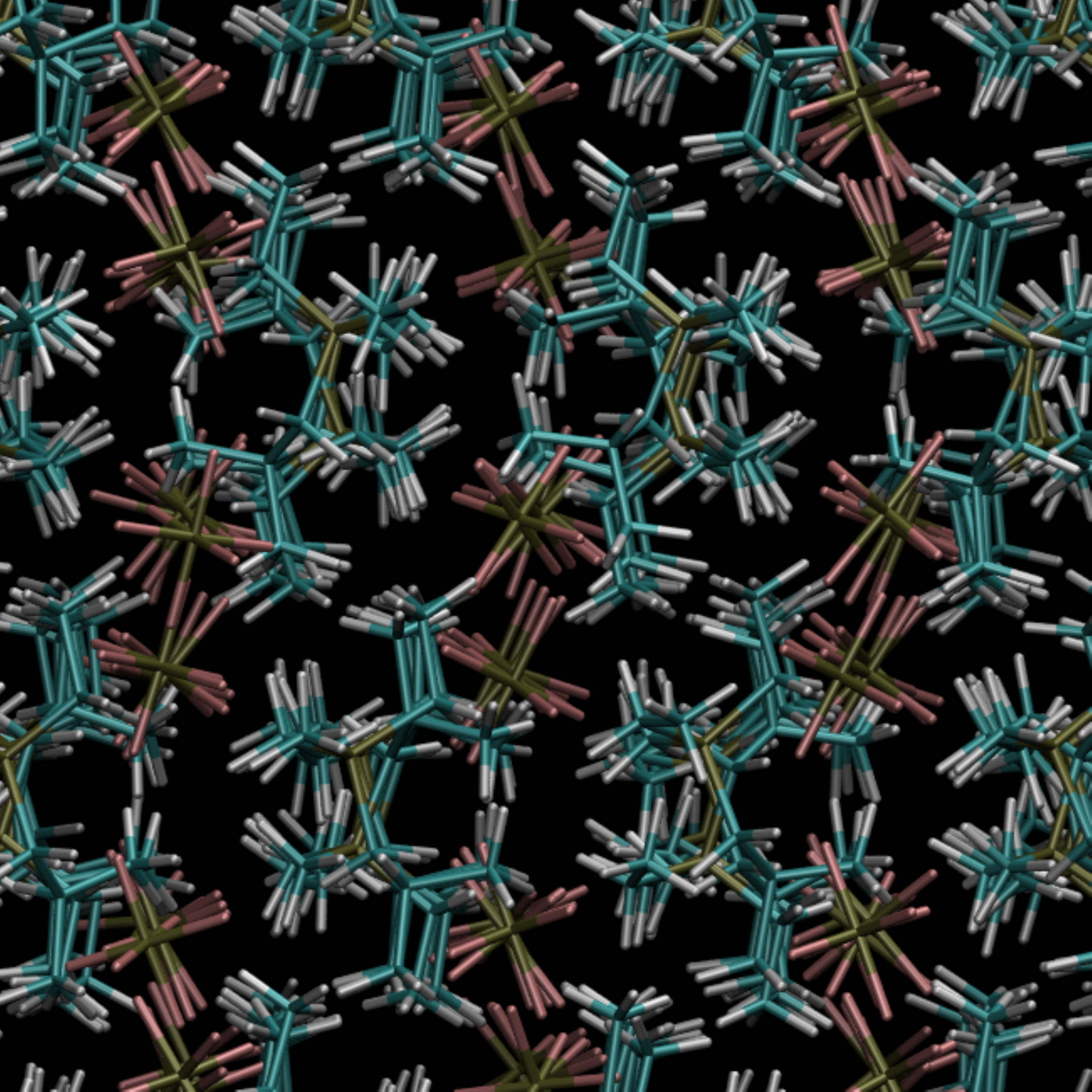}
\caption{Molecular structure of the [PF$_6$][P$_{1,2,2,4}$] dimer (top left). The other three panels are snapshots of a section of a well equilibrated configuration at 175 K, from the $x$ (top right), $y$ (bottom left) and $z$ (bottom right) directions of the simulation box. The different projections underline the anisotropy of the system.}
\label{views}
\end{figure}

We performed molecular dynamics simulations using the Gromacs v4.5.5 \cite{Hess:2008aa} simulation package. The initial conformation of the system was created from the low temperature (123 K) experimental crystalline structure, which corresponds to the $Pbca$ space group with orthorhombic parameters $a=12.9833$ \AA, $b=14.3871$ \AA\ and $c=15.19944$ \AA, and eight equivalent ion pairs in the unit cell \cite{Jin:2012aa}. All simulations were done under constant temperature and constant pressure conditions. The temperature was maintained at the reference value by the velocity rescaling algorithm \cite{Bussi:2007aa} with a time constant of 0.5 ps. An anisotropic Parrinello-Rahman pressure coupling was imposed with a reference pressure of 1 atm, compressibility of $4.5 \times 10^{-5}$ bar$^{-1}$ and time constant of 0.1 ps \cite{PARRINELLO:1981aa,NOSE:1983aa}. The dynamic equations of motion were integrated with the leap-frog algorithm using a time step 0.0005 ps. This relatively small time step was necessary in order to warrant the stability of the simulations. The simulation box contains 64 unit cells, or 512 ion pairs, which include 19,968 atoms. Tests with smaller system size produced excessive fluctuations and inconsistent behavior near the transition temperatures that were suppressed by using a simulation box with 512 ion pairs. It is important to note that finite size effects are important, and the results corresponding to 64 ion pairs are considerably different to the ones showed in this work.

A series of simulations were performed starting at 175 K from the perfect crystalline experimental structure. The temperature was then incremented by 25 K using as initial conformation the final structure (or a structure produced after 4 ns of simulation) of the system at the immediately lower temperature. This scheme was continued up to a final temperature of 500 K. When a qualitative change in the structure of the system was detected we performed extra simulations at intermediate temperatures. The total simulated time ($t_T$) is at least 6 ns for every case, with several temperatures reaching up to 22 ns. In general, simulations were continued until at least 4 ns of stabilized enthalpy and density were observed. Therefore, some simulations had to be extended for a considerable time due to an intrinsically slow kinetic behavior. For example, several intermediate temperatures were run for 20 ns to then discard the initial 10 ns since they show clear signs of drifting from the initial structure toward a different average state reached at the end of the simulation. The higher temperatures were run for 22 ns in order to have a good evaluation of the diffusion coefficients of both ions, which are quite small around the melting temperature. In order to limit the computational effort, we did not perform any free energy comparison to determine, from thermodynamic arguments, the relative phase stability. Although the thermodynamic analysis is very important, it is beyond the scope of this paper in which we take the simplified approach described above. In Table \ref{tabla-sims} we present a summary of all the simulations performed, along with the total simulated time $t_T$ and the time $t_P$ used to perform the analysis, discarding in all cases the $t_T-t_P$ initial interval.

\begin{table}[!b]
\caption{Summary of simulation runs. $t_T$ represents the total simulated time and $t_P$ the time used to perform time averages discarding the initial $t_T-t_P$ interval.}
\label{tabla-sims}
\begin{ruledtabular}
\begin{tabular}{lll||lll}
T (K)     &  $t_T$ (ns) & $t_P$ (ns)  & T (K)     &  $t_T$ (ns) & $t_P$ (ns)   \\ \hline
175       &  8   &  6   & 300       &  20  &  10  \\ 
180       &  6   &  4   & 305       &  20  &  10  \\ 
185       &  6   &  4   & 325       &  20  &  10  \\ 
190       &  6   &  4   & 330       &  20  &  10  \\
195       &  6   &  4   & 335       &  20  &  10  \\
200       &  8   &  5   & 340       &  20  &  10  \\
205       &  6   &  4   & 345       &  20  &  10  \\
225       &  6   &  4   & 350       &  20  &  10  \\
250       &  6   &  4   & 375       &  12  &  9   \\
275       &  6   &  4   & 400       &  13  &  5   \\  
280       &  6   &  4   & 425       &  22  &  15  \\  
285       &  6   &  4   & 450       &  22  &  15  \\  
290       &  6   &  4   & 475       &  22  &  15  \\  
295       &  20  &  12  & 500       &  14  &   9  \\  
\end{tabular}
\end{ruledtabular}
\end{table}

The interactions in the system were modeled using the CL\&P generic force field for ionic liquids, developed by Canongia-Lopes and P\'adua \cite{Lopes:2004aa,Lopes:2004ab,Canongia-Lopes:2006aa,Lopes:2008aa,Canongia-Lopes:2012aa}. The CL\&P force field is based on the OPLS-AA force field \cite{Jorgensen:1996aa}, from which it borrows the functional forms for all interaction types and several of the binding and non-binding parameters. Exclusive to the CL\&P are the partial charges associated to each atom. Using a MP2/cc-pVTZ(-f) {\em ab-initio} calculation as a reference to evaluate the electron density and corresponding electrostatic potential, the point charges for the model molecules are placed at each atom center and their value was determined by CHELPG methodology \cite{BRENEMAN:1990aa}. A downscaling of the ionic charges has been proposed by several authors to account for charge-transfer effects between the different ions \cite{Hunt:2006aa,Morrow:2002aa,Canongia-Lopes:2012aa}. In this work we initially study the system with no charge rescaling and every ion having a total charge of $\pm 1$. The results showed a shifted energy scale, in which all transition occurs at higher temperatures than the observed experimentally. A second series of simulations, which is presented here, were performed introducing a charge scaling factor $\alpha_q=0.8$. This number was chosen based on typical scaling values in studies of ionic liquids \cite{Liu:2011aa} and was not fitted to reproduce the experimental findings. The results obtained with the charge rescaling are in much better agreement with the available experimental data than those obtained without rescaling.

The exact functional forms of the force field used, along with the set of parameters are provided in the Supporting Information. An spherical cutoff of 12 \AA\ was imposed to the short range electrostatic and Lennard-Jones interactions, and the long range electrostatic contributions were accounted for using the particle mesh Ewald method \cite{ESSMANN:1995aa}.

\section{Results}
\label{results}

\begin{figure}[!t]
\includegraphics*[width=0.4\textwidth]{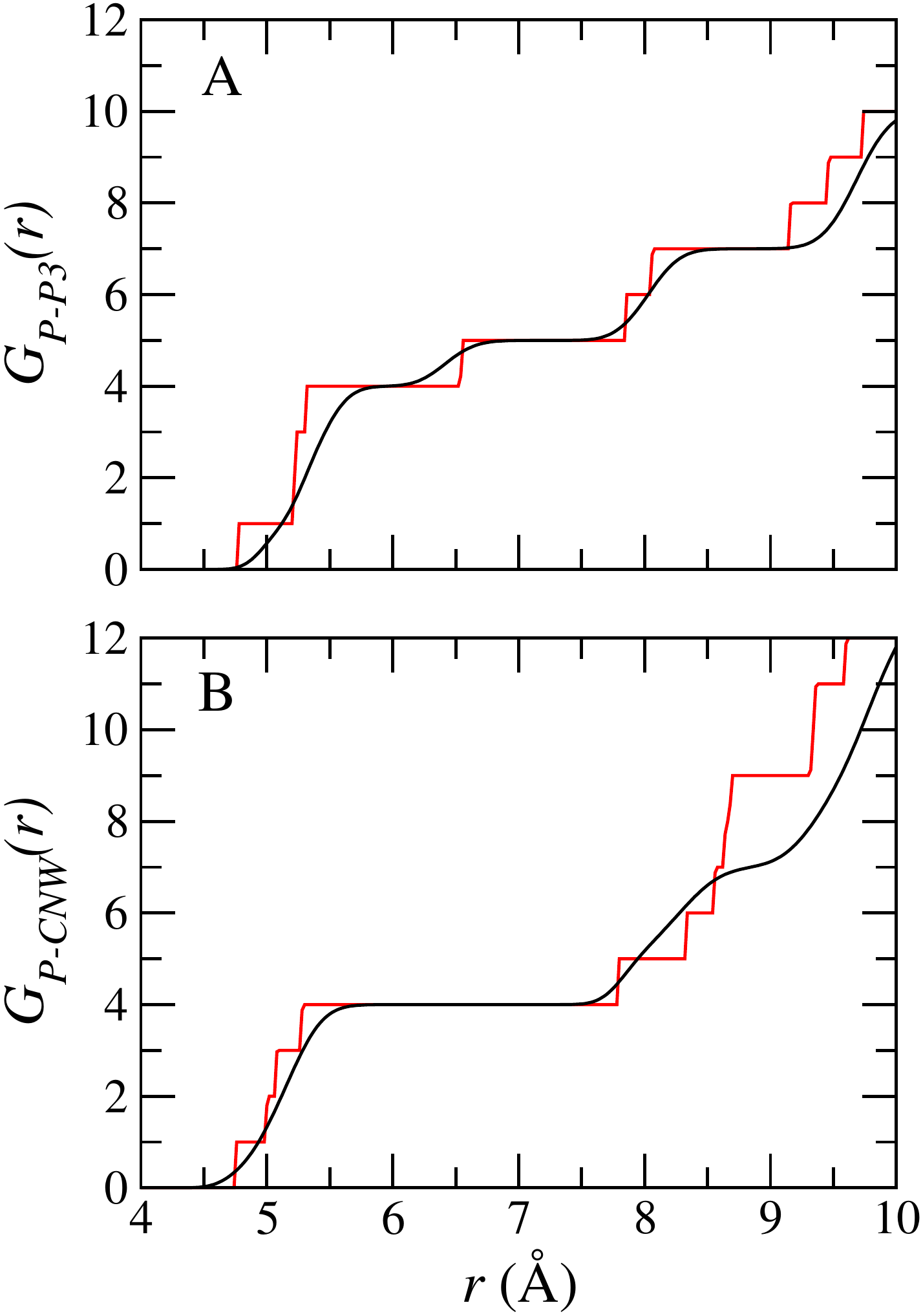}
\caption{Cumulative radial distribution functions for A) P-P3 and B) P-CNW atom pairs. The red lines correspond to the perfect crystallographic structure \cite{Jin:2012aa}, the black lines are simulation results at 175 K.}
\label{crdf}
\end{figure}

In Figure \ref{views} we show the molecular structure of [PF$_6$][P$_{1,2,2,4}]$ and three snapshots of the system to exemplify the clear anisotropic character of the crystalline structure. In particular, the projection on the $xz$ plane of the simulation box displays a hexagonal pattern where the [PF$_6$]$^-$ ions occupy channels defined by the larger [P$_{1,2,2,4}$]$^+$ ions. These snapshots correspond to the final configuration (8 ns) of the simulation at 175 K that was started from the perfect crystalline structure. The average structure obtained at this temperature after the initial relaxation allows us to compare the quality of the force field and simulation settings using the experimental crystalline data as reference.
The average lattice parameters, calculated over the last 6 ns of simulation at 175 K are $a=12.83$ \AA\, $b=14.94$ \AA\ and $c=15.64$ \AA, which all are within 3.8 \% of the experimental values that were measured at 123 K, showing that the distortion of the simulated system with respect to the experimental one is small. In order to further test the resemblance of the simulated crystal with experimental data we explore the local environment of the P atom of the hexafluorophosphate, through the cumulative radial distribution function $G(r)$ of the P3 and CNW atoms of the cations that are shown in Figure \ref{crdf}. Both simulated curves neatly smooth out the sharp perfect crystalline data, showing that P sees four first neighbors at at $r$(P-P3)$ \simeq 6$ \AA, and a total of seven neighbors at $r \simeq 8.5$ \AA. The four first neighbors are also clearly reflected by the $G(r)$ corresponding to the pair P-CNW. For crystalline systems we favor the cumulative radial distribution function over the standard pair distribution function, the reason being that even though the latter shows well defined peak positions it does not readily provide a quantitative vision of the number of first and second neighbors.

\begin{table}[!b]
\caption{Fitting parameters for $H$ vs. $T$ using Eq. (\ref{ecuafit}) for the two regions of interest.}
\label{tabla-fit}
\begin{ruledtabular}
\begin{tabular}{l|llllll}
Fit   &  $A_0$ & $A_1$   & $A_2$  & $A_3$   &  $A_4$ & $A_5$     \\ \hline
1  & 2.4746   &  197.15   & 2.16488  &  1.04322  &   -179.932 &  0.999862 \\ 
2  & 3.43888  &  332.926  & 11.2179  &  2.64094  &   -241.684 &  0.872201 \\
\end{tabular}
\end{ruledtabular}
\end{table}

\begin{figure}[!t]
\includegraphics*[width=0.4\textwidth]{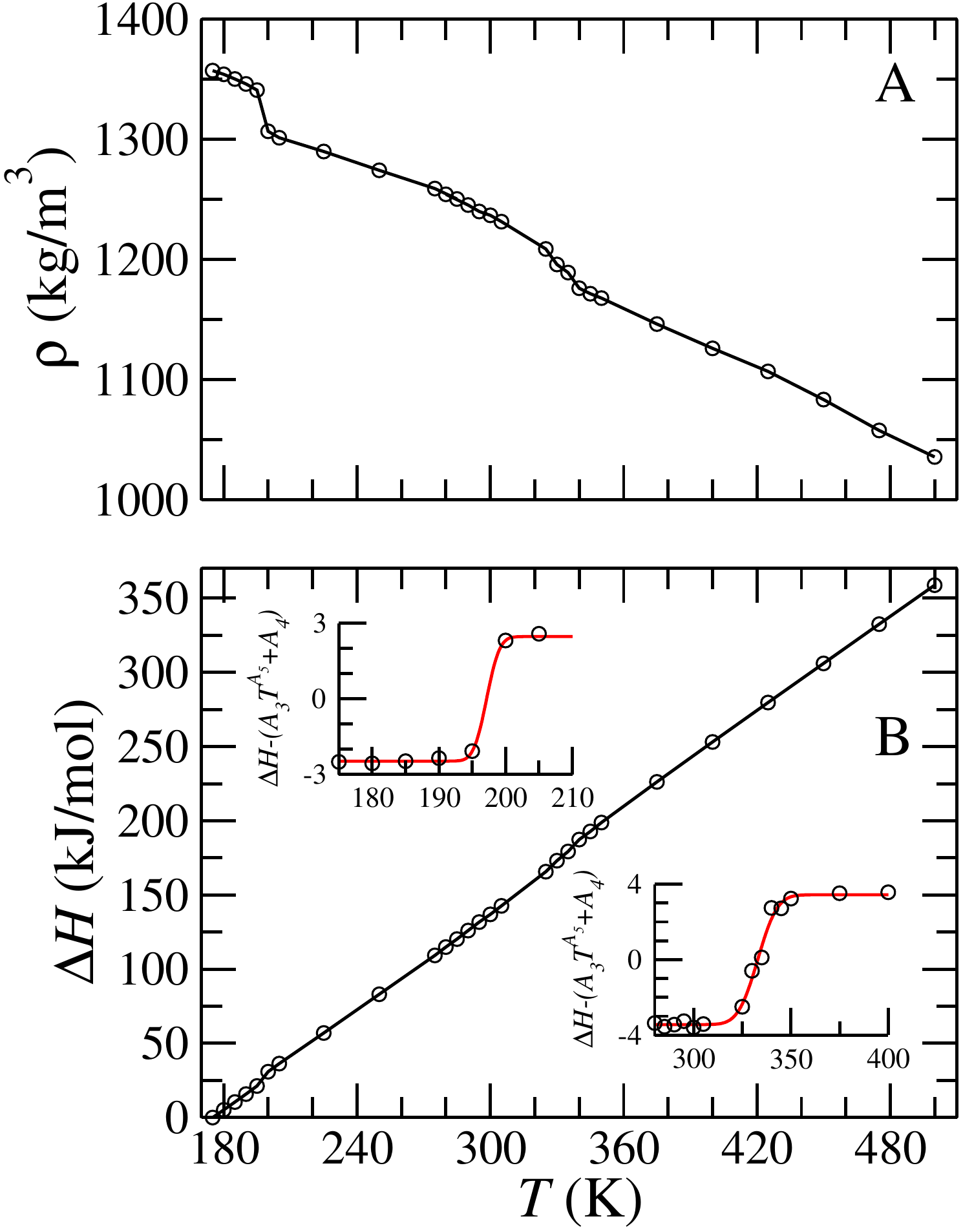}\\
\caption{A) Average density and B) enthalpy as a function of temperature. The inserts show $\Delta H - (A_3 T^{A_5}+A_4)$ for the simulation averages (symbols) and fit (red lines) using Eq. (\ref{ecuafit}). The enthalpy per ion pair is relative to the value obtained at 175 K.}
\label{density}
\end{figure}

The average system density and enthalpy as a function of temperature are displayed in Figures \ref{density}A and \ref{density}B, respectively. The density shows a sharp drop between 195 K and 200 K, a second broader change between 325 and 350 K. These features in the density are reflected in the enthalpy and therefore suggesting phase transitions occurring at those temperatures. In order to capture the enthalpy change we performed two fits of the $H$ vs. $T$ curve, the first for 175 K $\le T \le$ 275 K and the second for 280 K $\le T \le$ 475 K, using the fitting function 
\begin{equation}
f(T)=A_0 ~\text{erf}((T-A_1)/A_2)+A_3 T^{A_5}+A_4 ~~~~.
\label{ecuafit}
\end{equation}
Both fits achieve a correlation coefficient better than 0.99999, and the fitting parameters are summarized in Table \ref{tabla-fit}. The inserts on Figure \ref{density}B show the enthalpy minus the leading terms from the fits, i.e. $\Delta H - (A_3T^{A_5}+A_4)$. The transition temperatures are essentially the parameter $A_1$ from the enthalpy fits, i.e., 197 K and 333 K for the first and second transition, respectively. The calculated enthalpy changes are 5 kJ/mol and 7 kJ/mol for the first and second transition, respectively. Then we can estimate the corresponding entropy change for those transitions, resulting in 25 J/K and 21 J/K, respectively. These values are in line with typical values observed in transformations involving plastic phases in general \cite{Sherwood:1979aa} and for this system in particular \cite{Jin:2012aa}.

The molecular rotations are characterized using rotational self correlation functions $C_\delta(t)=\langle \vec{C}_\delta(t_0+t).\vec{C}_\delta(t_0) \rangle$. Here, $\delta$ refers either to {\it i}) a group of three atoms with $\vec{C}_\delta$ being the unit vector normal to the plane defined by those atoms or {\it ii}) a group of two atoms defining a chemical bond and in this case $\vec{C}_\delta$ is a unit vector along that bond. A unit vector rotating over the complete sphere will relax to zero in a characteristic time. However, if the rotation spans only a defined solid angle then $C_\delta(t)$ will reach an asymptotic value larger than zero. For example, the self correlation function of a vector spinning on a right regular cone of aperture $2\theta$ will relax to $\cos \theta$ and not to zero as it would if the vector could span a whole sphere. Having this in mind, we analyze the $C_\delta(t)$ curves obtained from the simulated trajectories by fitting a function of the form:
\begin{equation}
C(t)=C_0 e^{-(t/\tau)^\beta} + \alpha.
\label{stretch}
\end{equation}
The constant $C_0$ is included to account for the fast (a few ps) relaxation of the molecular librations. The additive constant $\alpha$ is to capture the asymptotic limit of $C(t)$. The relaxation time is calculated as $\tau^\ast=\Gamma(1/\beta)\,\tau/\beta$.

\begin{figure}[!b]
\includegraphics*[width=0.4\textwidth]{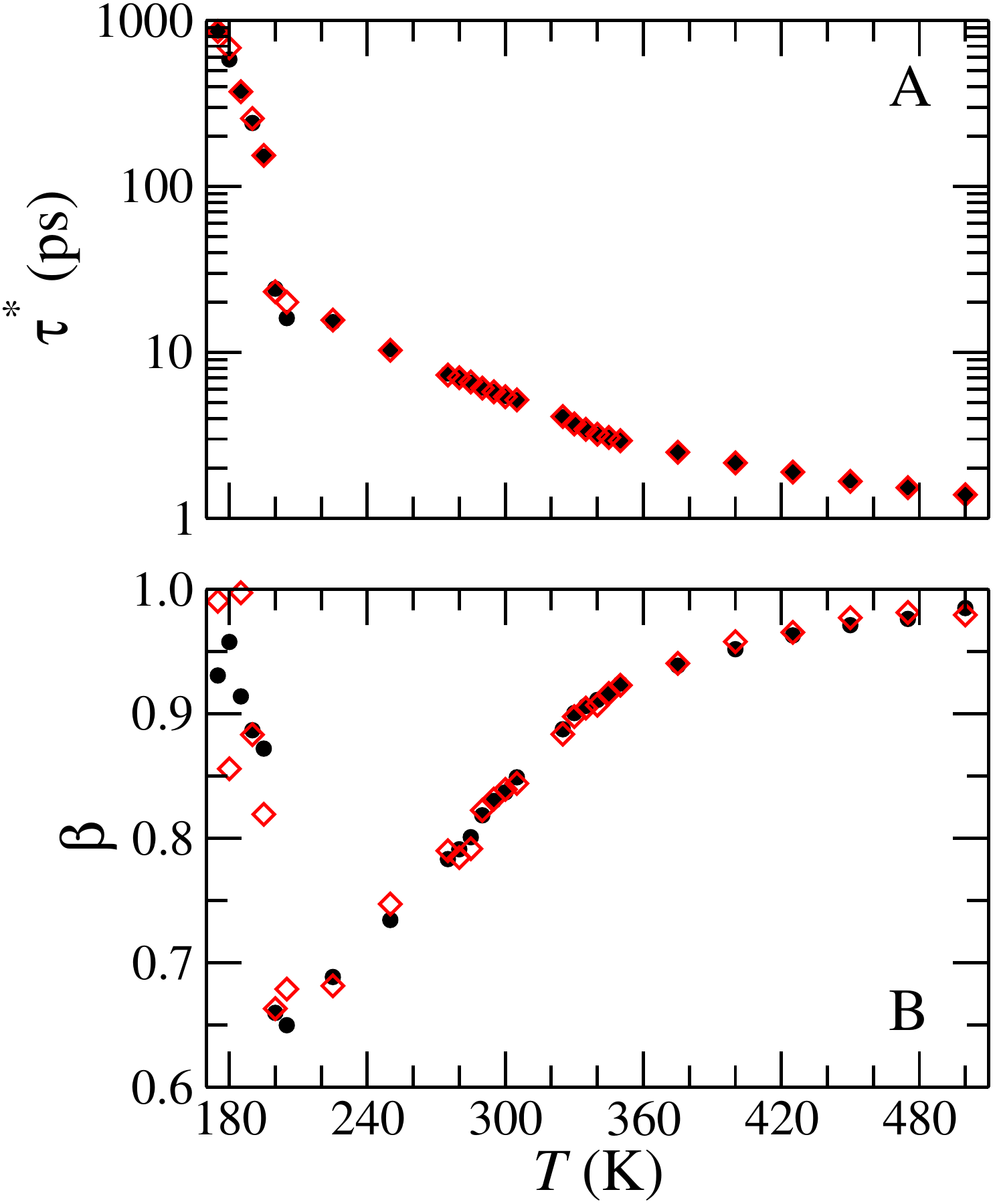}
\caption{A) Relaxation time and B) stretching exponent for the rotations of [PF$_6$]$^-$ as a function of temperature. Black circles represents the results from three F atoms defining the direction of the relaxing vector. Red diamonds are the results using the molecular PF bonds.}
\label{taupf6}
\end{figure}

The orthorhombic crystalline structure at which the system was prepared remains stable up to a temperature of 197 K. The phase at temperatures immediately above 197 K corresponds to a {\em semi} plastic phase, where the PF$_6$ ion rotates freely. Indeed, in Figure \ref{taupf6}A we show the relaxation times for the rotations of the anion as a function of the temperature, showing a clear breaking at the transition temperature. The two sets of relaxation times shown in Figure \ref{taupf6}A, which contain essentially the same information, are originated from the calculation of the rotational self correlation function of a vector defined as the normal to the plane defined by three F atoms (black circles), or from the vectors defined by the PF bonds (red diamonds). The fit of the results using a stretched exponential function, Eq. (\ref{stretch}), is very good with a correlation coefficient higher than 0.99 in every case. The asymptotic value $\alpha$ is statistically zero for all cases. In Figure \ref{taupf6}B we plot the stretching parameter $\beta$ of the fit as a function of temperature. Interestingly, this plot also shows a distinct change in behavior at the transition temperature. A single exponential relaxation does not capture the anion rotational relaxation. We interpret this result in terms of the highly anisotropic environment surrounding the anions, as shown in Figure \ref{views}. Rotation about each one of the three main axes of PF$_6$ ions are affected in a different way and therefore the combination of all rotations cannot be described by a single exponential curve. The sudden change of the stretching parameter $\beta$ at the transition temperature reflects the modification of the environment surrounding the anions in the warmer phase. As the temperature increases beyond 197 K, $\beta$ monotonically increases approaching the exponential behavior as the system becomes more homogeneous. The equivalence of these two sets of measurements indicates that the molecular rotations are uncorrelated with the physical molecular bonds. 

The onset of the {\em semi} plastic phase in the experimental work was not detected, and Ref. \cite{Jin:2012aa} states that for $T =$ 193 K the system is not rigid and some motion is present, in particular for the anions, due to negligible steric hindrance to the rotations of [PF$_6$]$^-$. The experimental {\em semi} plastic phase (phase IV) extends up to 298 K. Phase IV is characterized by fast thumbing of the anion. These are indeed the characteristics of the model above 197 K as revealed by the rotation of the anion presented in Figure \ref{taupf6}. Therefore, the model predictions are in line with the experimental results for the lower temperature limit of phase IV.

\begin{figure}[!t]
\includegraphics*[width=0.5\textwidth]{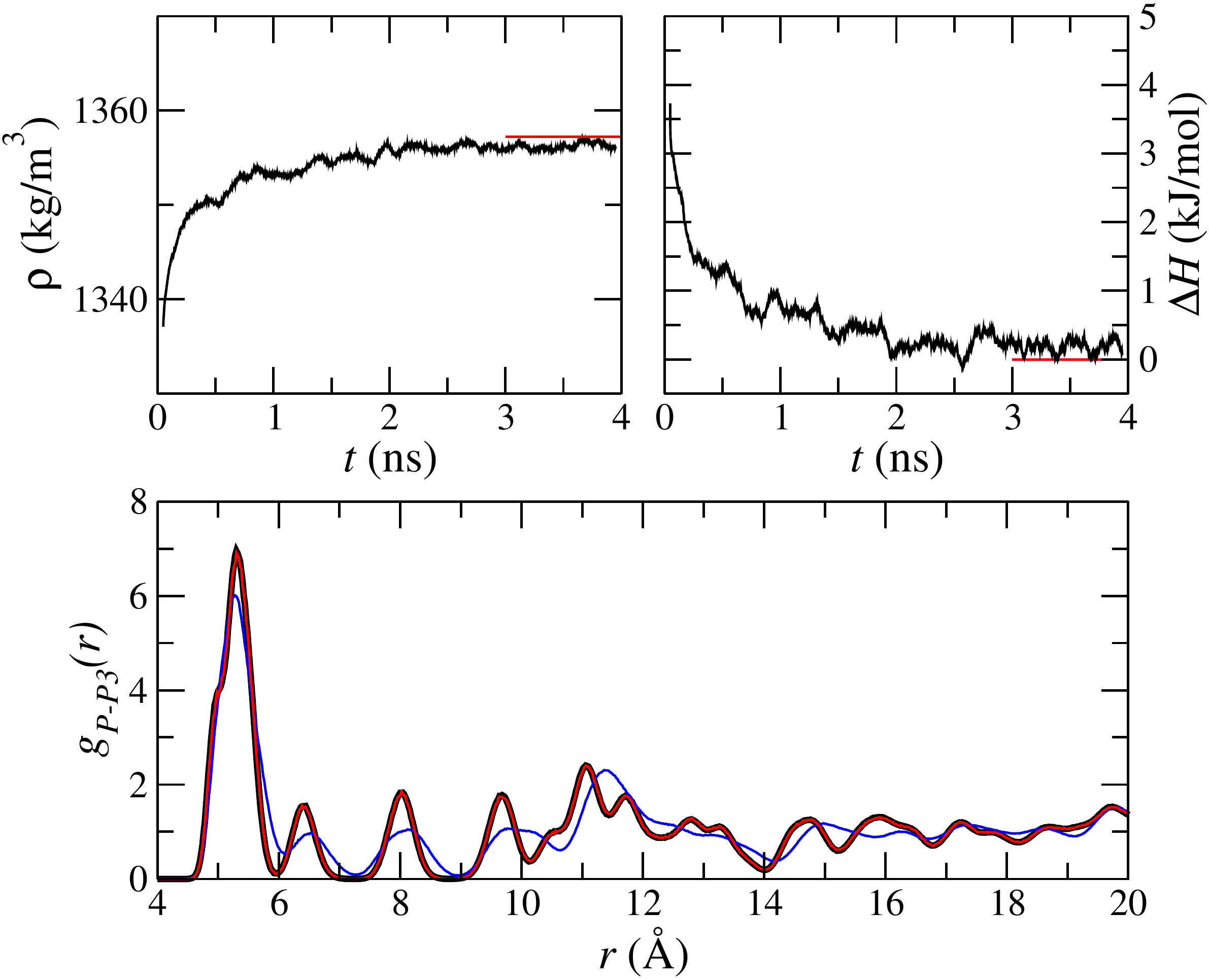}
\caption{Refreeze at 175 K from the {\em semi} plastic phase obtained at 200 K. The top two panels show the time evolution of the density (left) and enthalpy of the system (right). The bottom panel shows the P-P3 radial distribution functions for a well equilibrated system at 175 K (black line),  at the start of the simulation  corresponding to the equilibrium structure at 200 K (blue line) and  after 4 ns of simulation when the system refroze to the crystalline structure (red line).}
\label{refreeze}
\end{figure}

In order to check the stability of the crystalline and {\em semi} plastic phases around 197 K we performed a simulation at 175 K, starting from a configuration corresponding to the {\em semi} plastic phase obtained at 200 K. On Figure \ref{refreeze} we show the time evolution of the density and enthalpy of the system, and a comparison of the radial distribution function between P and P3 at the end of the simulation together with the corresponding curves obtained from the initially crystalline system simulated at 175 K. The system clearly returns to the crystalline phase in a time of approximately 4 ns. This is indeed an interesting and remarkable outcome. Typically, MD simulations can describe systems undergoing a phase transformation to a state of higher entropy but lower free energy at a given temperature. The inverse transformation is more difficult to capture. In this case, however, the fact that the overall structure of the system remains with crystalline order facilitates for the system the finding of a path towards the lower entropy state upon the decrease of the temperature. The magnitude of the reduction of the temperature, which is the driving force for the phase change, has to be large enough so that the phase change is observable in MD times. In this case, we did not observe convincing signs of recrystallization in a similar simulation at 180 K. 

\begin{figure}[!t]
\includegraphics*[width=0.4\textwidth]{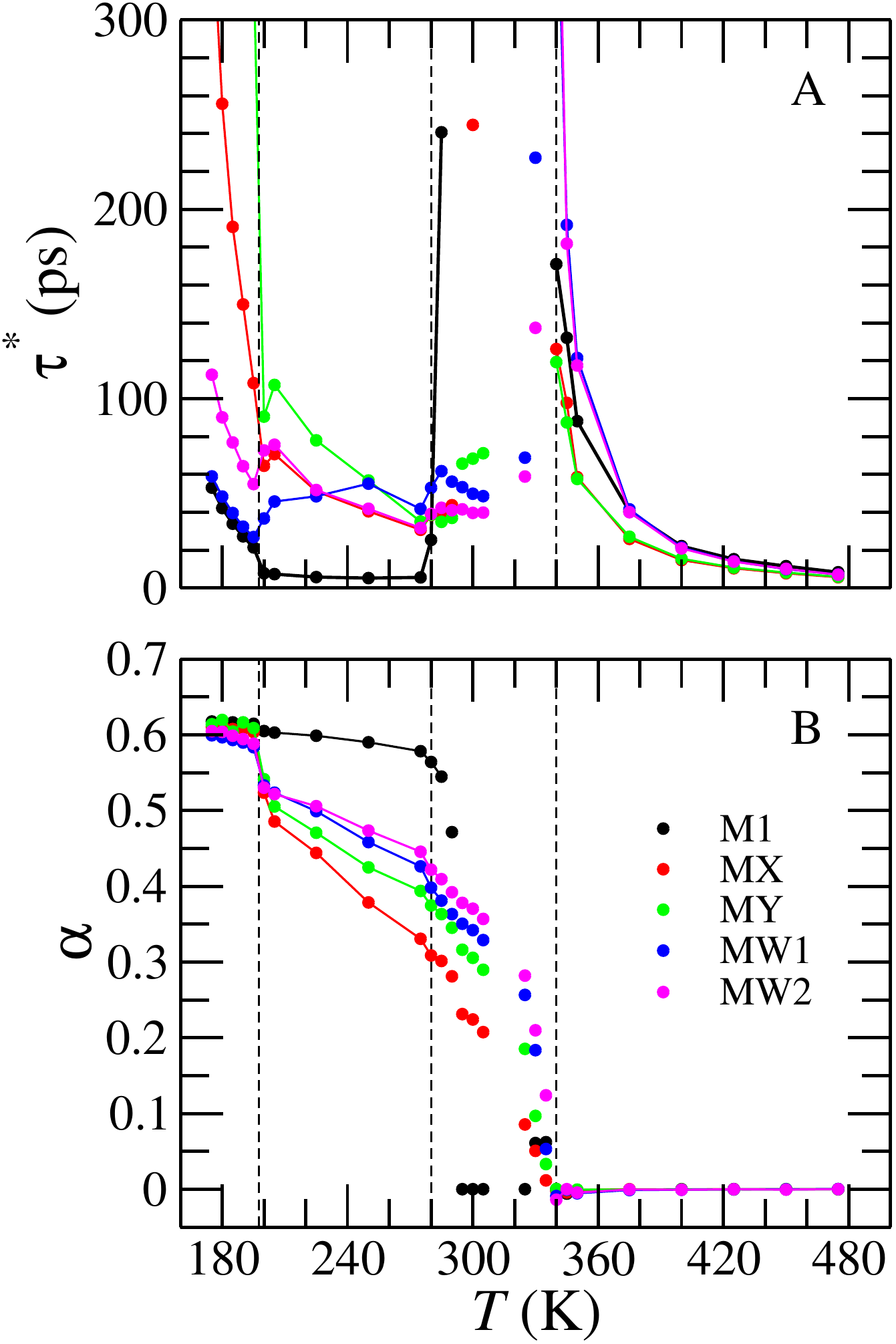}
\caption{A) Relaxation time and B) asymptotic value of the self correlation function of the five methyl groups of the P[1,2,2,4] cation. The dashed vertical lines are placed at the transition temperatures 197 K, 280 K, and 340 K. The connecting lines were dropped for 280 K $< T < 340$ K due to the noise in the data, especcially for $\tau^\ast$, originated by the slow rotations of the cation.}
\label{methyls}
\end{figure}

The analysis of the rotational dynamics of the cation is more complex than the anion because of its larger size and intramolecular flexibility. Nevertheless, it is possible to draw a general picture by looking at the rotational autocorrelation functions defined by different groups of atoms. The crystalline structure obtained after 8 ns of simulation at 175 K display a strong stability with no overall rotations of the cation. For all groups $\delta$ composed by any combination of three heavy adjacent atoms of the cation backbone we observe only librational motions at 175 K. However all the methyl groups display slow rotations about the axis defined by the corresponding C atom. The relaxation time is different for all the different methyl groups, even for those that are equivalent within the molecule like MX and MY on one hand, and MW1 and MW2 on the other hand. Considering the effective rigidity of the cation at low temperature, then the different relaxation times between methyl groups has to be related to their local environment.

\begin{figure}[!b]
\includegraphics*[width=0.4\textwidth]{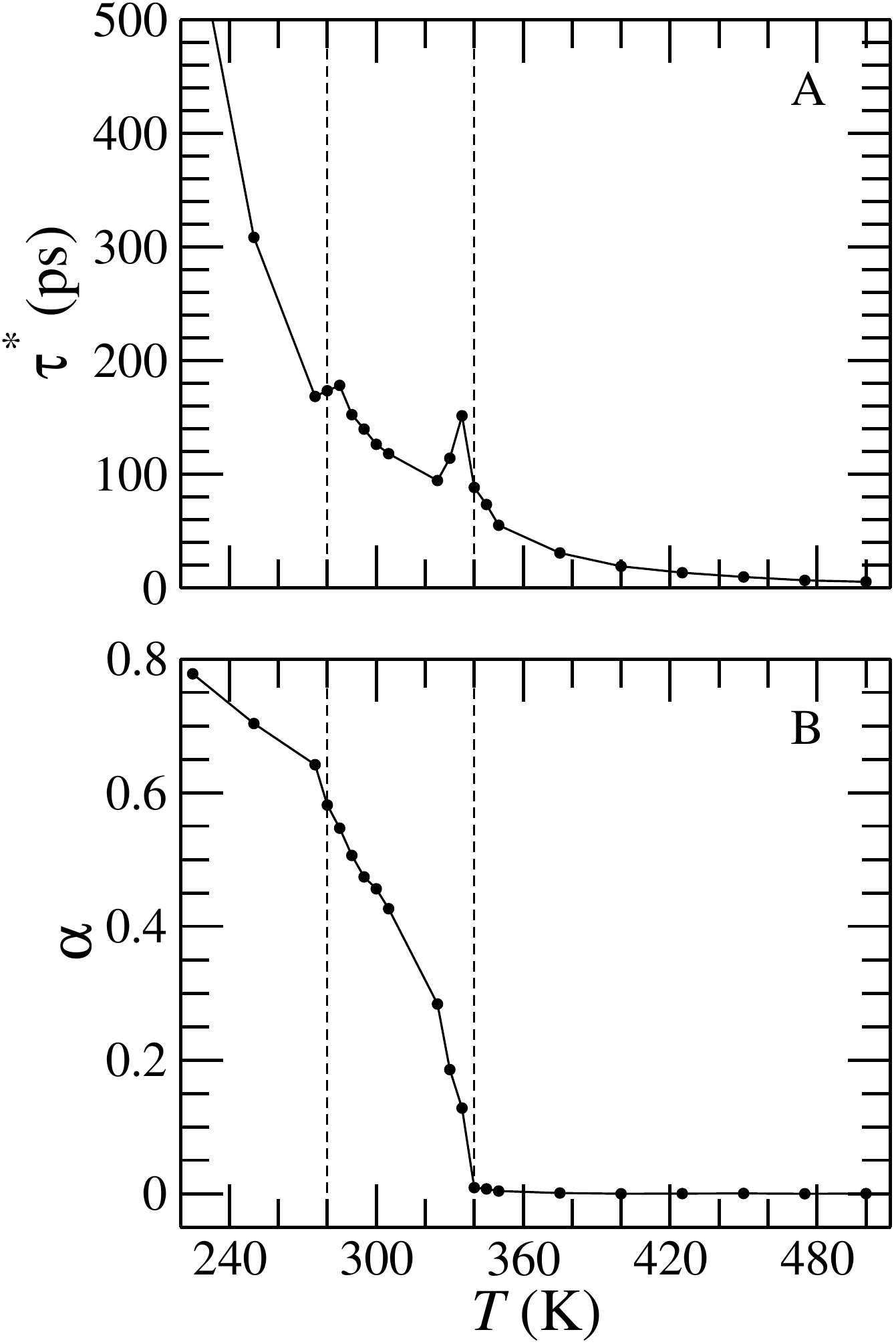}
\caption{A) Relaxation time and B) asymptotic value of the self correlation function of the group $\delta_{CNW}=\{CNW,CW1,CW2\}$. The dashed vertical lines are placed at the transition temperatures 280 K and 340 K.}
\label{cfar}
\end{figure}

\begin{figure*}[!ht]
\begin{center}
\includegraphics[width=0.9\textwidth]{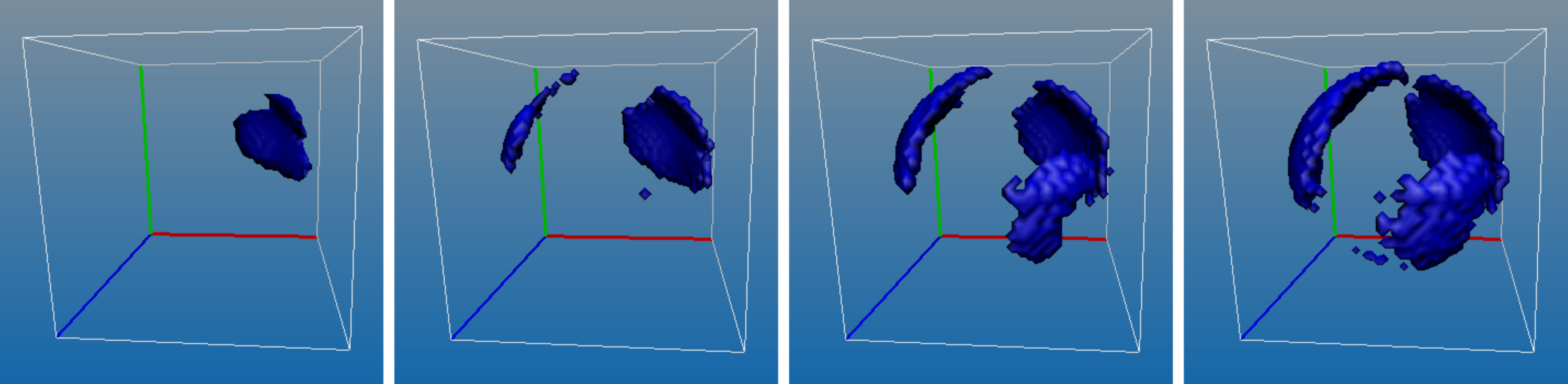}  
\caption{Density isosurface of the CP1 atom, relative to the position of the central P3 atom, as a function of temperature. From left to right the densities correspond to 225 K, 275 K, 300 K and 325 K. For 340 K the corresponding figures shows a closed sphere.}
\label{fotos}
\end{center}
\end{figure*}

An interesting picture emerges from looking at the temperature dependence of the methyl relaxation. In Figures \ref{methyls}A and \ref{methyls}B we show the relaxation time $\tau^\ast$ for the five methyl groups and the asymptotic value $\alpha$ corresponding to the fits of the rotational autocorrelation functions as a function of the temperature. There are four distinctive regions divided by $197$ K, $280$ K and $340$ K. For $T<197$ K all the groups are rotating about their corresponding axis, defined by P3-CP1, CPX-CEX, CPY-CEY, CNW-CW1 and CNW-CW2, which remain essentially fixed in space. This is clear from the asymptotic values $\alpha \sim 0.6 \sim \cos (\theta_{HCH}/2)$. The relaxation time $\tau^\ast$ monotonically decreases for $T <$ 197 K. For $T$ immediately above 197 K we observe a sudden drop in $\alpha$, in particular for MX and MY and to a lesser extent for MW1 and MW2. A clear discontinuity is also observed in $\tau^\ast$ at 197 K. This pattern is consistent with the structural transformation indicated by the decrease in the system density at that temperature. Between 197 K and 280 K both parameters smoothly decrease with increasing temperature, except for the measured $\tau^\ast(MW1)$ that remain stable or slightly increase. The decrease in $\alpha$ indicates an increase in the amplitude of the librations experienced by the C atoms. Between $280$ K and $340$ K there is a dramatic increase in the relaxation times that goes beyond the time scale of the simulations and therefore our estimations are not reliable. This increase in $\tau^\ast$ occurs at 280 K for M1 and MX, but at 310 K for the other methyl groups. The asymptotic value $\alpha$, although displaying a degree of noise, keep decreasing as $T$ approaches $340$ K. This trend suggests the sequential activation of rotational modes that could be the result of the intramolecular flexibility of the cation and/or an overall molecular rotation. For $T>340$ K the asymptotic value $\alpha$ vanishes in all cases and $\tau^\ast$ decreases monotonically with increasing temperature.

In order to explore the origin of the long methyl relaxation times for 280 K $< T <$ 340 K we investigate the self correlation function of the vector defined as the normal to the plane defined by the atoms CNW, CW1 and CW2. In Figure \ref{cfar} we show $\tau^\ast$ and $\alpha$ as a function of temperature. These curves show distinctive features around the transition temperatures, but no dramatic increment in the relaxation times. This behavior can be explained if the axis of rotation for the this group of atoms is nearly parallel to the slow rotational modes that are responsible for the large increase in the relaxation times of the methyl groups for 280 K $< T <$ 340 K.

We turn now our attention to the innermost region of the molecule by monitoring the rotations of the four bonds involving the P3 atom. As expected, we see in the rotational relaxation functions $C(t)$ (see Supplemental Information) the three regimes revealed by the methyl groups. There is however a slightly different behavior of the bond P3-CPW with respect to the other three: P3-CP1, P3-CPX and P3-CPY. For $T < 280$ K the relaxation of all groups is relatively fast to a finite asymptotic value $\alpha$. For $T>340$ the relaxation of all groups is very fast to a value $\alpha=0$. For the intermediate region the relaxation times are longer than the timescale of the simulations for all bonds except the P3-CPW that extends the fast relaxation mode up to 305 K. This behavior is compatible with the activation of rotational modes at 280 K aligned with the P3-CPW bond, which in turn is parallel the longitudinal axis of the cation. For temperatures higher than 305 K the [P$_{1,2,2,4}$] molecule starts to undergo slow rotations along the other directions and the rotational relaxation of the P3-CPW bond display the corresponding long relaxation times. It is worth to emphasize that at the intermediate temperature regime the kinetics of the system is dominated by the longitudinal rotation of the cation, which is intrinsically slow for temperatures just above 280 K and speeds up towards 340 K. This directly affects the time required for the system to approach equilibrium, and therefore long simulation times are needed.

\begin{figure*}[!ht]
\begin{center}
\includegraphics[width=0.9\textwidth]{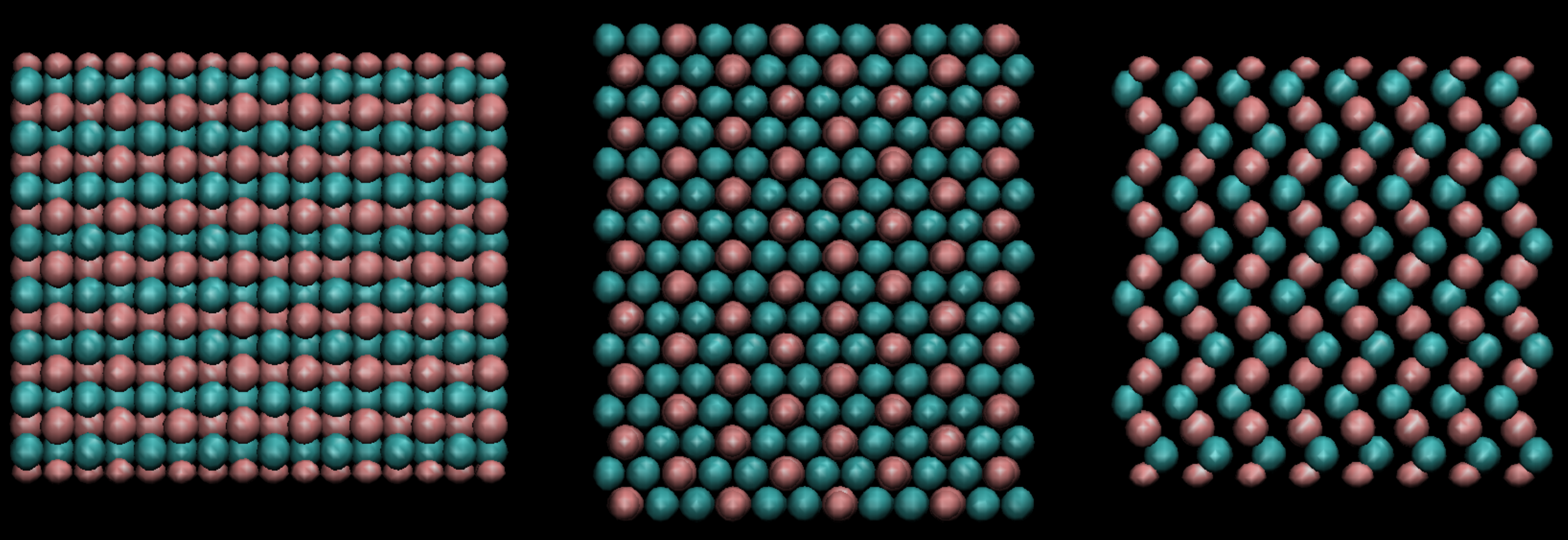}  
\caption{Average density of the P atoms corresponding to the PF$_6$ (pink) and P$_{1,2,2,4}$ (cyan) calculated in the {\em full} plastic phase at 425 K. From left to right, the panels are the normal to $x$, $y$ and $z$ directions of the simulation box, respectively.}
\label{densfull}
\end{center}
\end{figure*}

A clear picture of the kinetic behavior in this intermediate temperature regime emerges from looking at the 3D densities of the position of the CP1 atoms, relative to the central P3 atom of the same molecule. In order to make the proper system average, we separate the 512 molecules in eight groups according to their relative place in the unit cell (see Supplemental Information). At 225 K and lower temperatures, the densities reveal a single spot that imply no rotations of CP1 around the P3. At 275 K the densities show two separate spots due to sudden rotations, or jumps, of approximately 120$^\circ$ between two locally stable configurations. Increasing the temperature to 300 K and 325 K results in a clear enhancement of this process, that now reach the three equilibrium positions separated by 120$^\circ$. At 340 K and higher temperatures the corresponding plots cover a whole sphere indicating that the rotations are not anymore restricted to the local equilibrium positions determined by the intramolecular configurations. All these results indicate that the model systems enters into the {\em full} plastic phase at 340 K. In Figure \ref{densfull} we show the structure of this phase s displayed by the density of occupancy of the P atoms of two ions, calculated at 425 K. The projections show the charge ordering and anisotropy of the structure, that preserves the hexagonal channels of cations along the $y$ axis of the simulation box, enclosing the anions in a neat hexagonal pattern.

The onset of longitudinal rotations of the cation that we found at 280 K corresponds to the onset of the phase III at 298 K reported in the experimental study of Jin et al. \cite{Jin:2012aa}. The transformation to a {\em full} plastic phase at 340 K also agrees well the experimental determination of the transformation to phase II at 343 K. Our model system does not show two distinguishable signals in the enthalpy, which continuously evolves to display a change at the intermediate temperature 333 K. However the kinetic behavior of the system provides clear signatures of the two transformations. The experimental work reports yet another transformation to phase I at 393 K and melting at 435 K. The difference between phase II and phase I is essentially  in the translational kinetics of the ionic components: while only the anion can diffuse in phase II both ions have a diffusive dynamics at the highest solid phase. Then we evaluated the diffusive coefficient of the components of our model by calculating the mean squared displacement of the center of mass of each ion, for the different temperatures. The results are presented in Table \ref{dif}. While the smaller anion shows signs of diffusion at 400 K, the cation display a slow diffusive mode at 450 K. We interpret this onset of cation diffusivity as a signature for the melting of the system. It is not clear to us how to differentiate in a small system such as the simulated in this work, a solid with significant diffusion in all its components from the liquid state.

\begin{table}[!b]
\caption{Diffusion coefficients calculated from $\Delta r^2$ vs. $t$ curves.}
\label{dif}
\begin{ruledtabular}
\begin{tabular}{l|llllll}
       &      D (10$^{-5}$cm$^2$/s)  \\ \hline
T (K)  &  PF$_6$  & P$_{1,2,2,4}$    \\ \hline
400    &   0.0002 & 0.0000 \\ 
425    &   0.0014 & 0.0000 \\
450    &   0.0051 & 0.0001 \\
475    &   0.0104 & 0.0002 \\
500    &   0.0285 & 0.0019 \\
\end{tabular}
\end{ruledtabular}
\end{table}

\begin{figure}[!t]
\includegraphics*[width=0.45\textwidth]{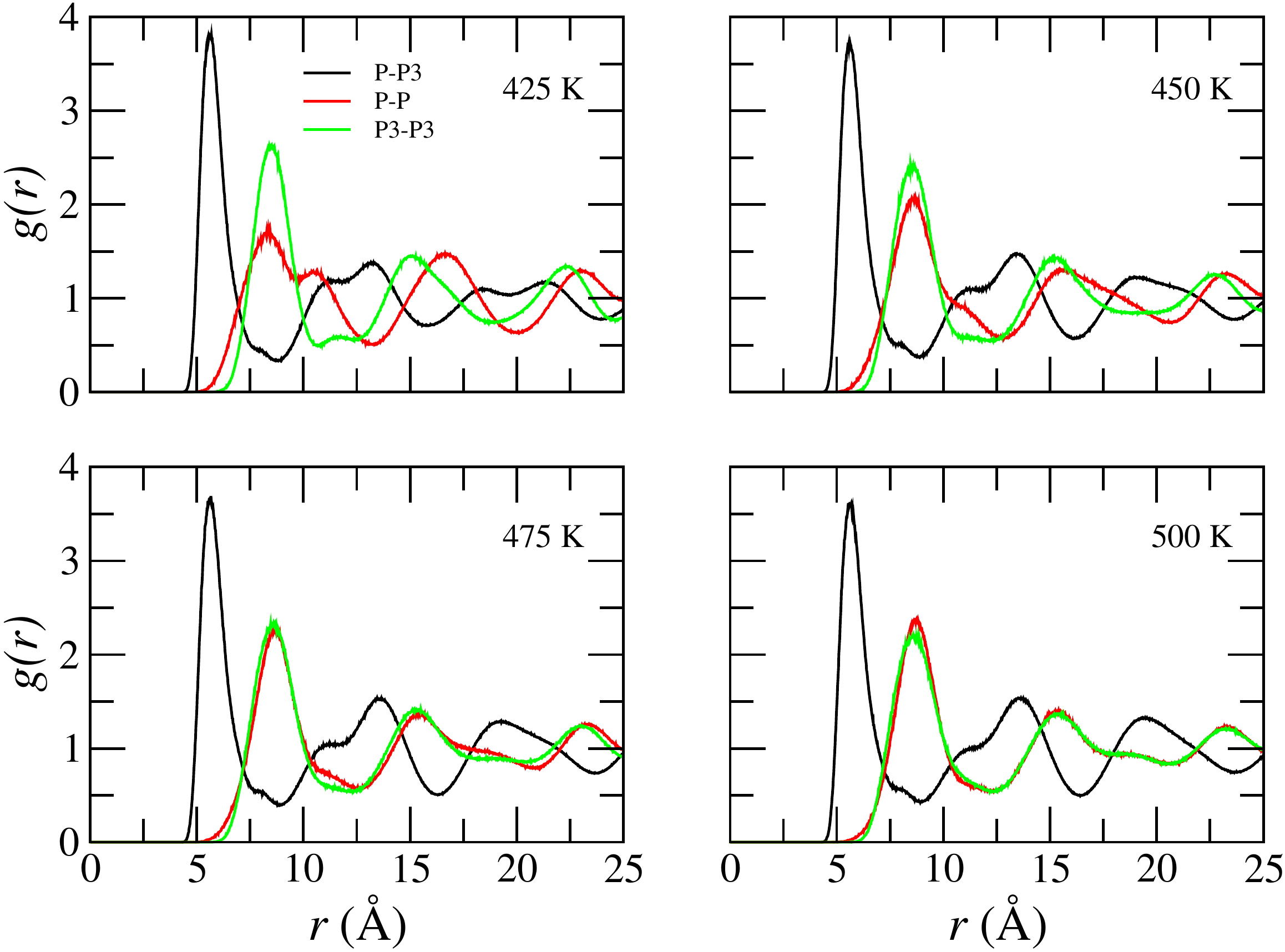}
\caption{Pair distribution functions for all combination of P atoms.}
\label{gs}
\end{figure}

The structural changes occurring at high temperature are revealed by the pair distribution functions, $g(r)$. For example, in Figure \ref{gs} we show the $g(r)$ for the P-P, P3-P3 and P-P3 pairs at 425 K, 450 K, 475 K and 500 K. At 425 K there is a clear difference between the P-P and P3-P3 pair distribution function. However, as the temperature increases the two functions overlap, indicating that even though the two ions have quite different size they effectively occupy the same volume in the liquid state.

\section{Discussion}
\label{final}

The present study of [PF$_6$][P$_{1,2,2,4}$] is based on the CL\&P force field, which was used with a charge rescaling $\alpha_q=0.8$. The results show that the model systems transforms from a crystalline orthorhombic phase at low temperature to a {\em semi} plastic phase at 197 K and to a {\em full} plastic phase at 340 K. An intermediate regime is identified at 280 K and higher temperatures that ends as the system transforms to the {\em full} plastic phase. This intermediate regime is characterized by a sporadic, jump-like rotations of the cation. These rotations are, for 180 K $\le T <$ 325 K, along the longitudinal axis of the cation and become isotropic at 325 K. The system melts at $\sim$ 450 K to a liquid state where the two ionic species effectively occupy the same volume.

The prediction of the model are qualitatively in agreement with the recent experimental work of Jin et al. \cite{Jin:2012aa}. The overall picture unveiled from the series of experiments used in Ref. \cite{Jin:2012aa} is reproduced here with a simple, pair-additive, atomistic model. Although more sophisticated approaches are being considered, such as models with atomic polarizations \cite{Borodin:2003aa}, the availability of a sufficiently good simple pair-additive model will allow the study larger systems including interfaces that are a very important factor determining the performance of many devices. To achieve quantitative agreement of an atomistic model with the experimental data, and therefore to be able to claim to have quantitative predictive power, is the ultimate goal for the simulation community. Nevertheless, this is a difficult objective and its achievement requires consistent efforts. This work was though, from the very beginning, as an effort towards the goal.

\section{acknowledgments}
The author acknowledges Dr. Pablo Serra from Fa.M.A.F., Universidad Nacional de C\'ordoba (Argentina), for enlightening discussions during the preparation of this paper.

\section{Supporting Information Available}
Detailed description of the force-field parameters, examples of the cation rotational autocorrelation functions and the corresponding analysis method is presented in a separate file. This information is available free of charge via the Internet at http://pubs.acs.org.

%
%

\pagebreak

\begin{center}
\large{\bf{\underline{Supplementary Information}}}
\end{center}

\vspace{1.8cm}

\section*{CL\&P Force Field for [PF$_6$][P$_{1,2,2,4}$]}

\begin{figure}[!t]
\includegraphics[width=0.78\textwidth]{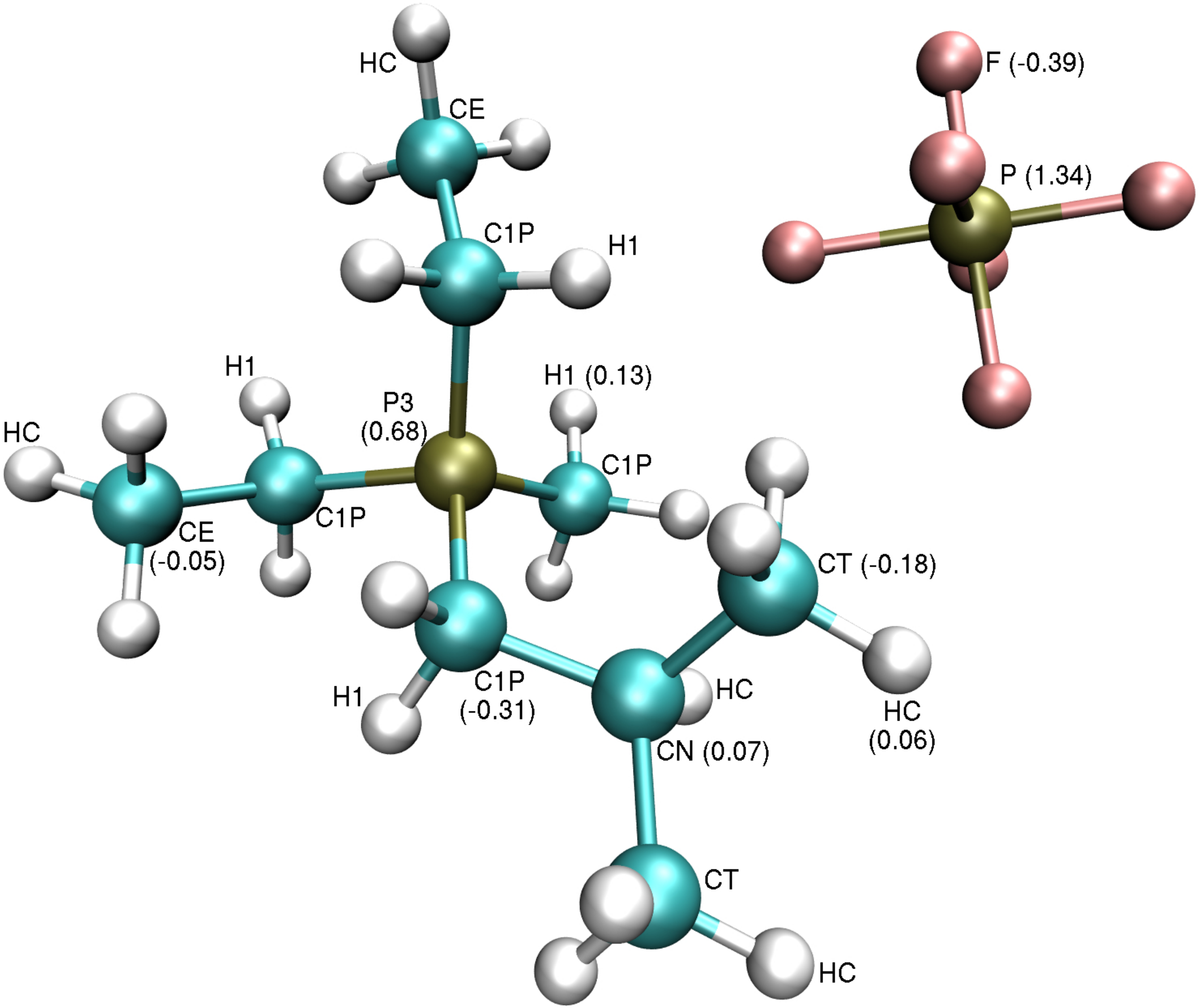} \vspace{1cm} \\
\caption{Detailed structure of the  [PF$_6$][P$_{1,2,2,4}$] dimer including the CL\&P atom types and partial charges.}
\label{dimerq}
\end{figure}

\noindent
The CL\&P force field, developed by Canongia Lopes and P\'adua and extensively described in a series of papers (Refs. 14-18 of main paper) takes the functional form of the well known OPLS-AA (Ref. 19 of main paper) force field:
\begin{eqnarray}
U&=&\sum_{ij}^{N_b}\frac{k_{r,ij}}{2}\left (r_{ij}-r_{0,ij} \right )^2 + \sum_{ijk}^{N_a}\frac{k_{\theta,ijk}}{2}\left (\theta_{ijk}-\theta_{0,ijk} \right )^2  \nonumber \\
&& + \sum_{ijkl}^{N_d} \sum_{m=1}^4 \frac{V_{m,ijkl}}{2} \left [ 1 + (-1)^m \cos (m \phi_{ijkl}) \right ] \\
&& + \sum_{i}^N \sum_{j \neq i}^N \left \{ 4 \epsilon_{ij} \left [ \left ( \frac{\sigma_{ij}}{r_{ij}}
\right )^{12} - \left ( \frac{\sigma_{ij}}{r_{ij}} \right )^{6}  \right ]
+  \frac{\alpha_q}{4 \pi \epsilon_0} \frac{q_i q_j}{r_{ij}}
\right \} \nonumber
 \end{eqnarray}
where $N$, $N_b$, $N_a$ and $N_d$ are the number of atoms, bonds, angles and dihedrals in the system, respectively.
The first three terms are the bonding interactions, and the last two terms are the non-bonding interactions. Harmonic bonds between atoms $i$ and $j$ require an equilibrium bond length $r_{0,ij}$ and force constant $k_{r,ij}$. Similarly, the harmonic angles are characterized by equilibrium angle $\theta_{0,ijk}$ and the corresponding force constant $k_{\theta,ijk}$.
The dihedral interactions are described by four cosine terms as a Fourier series with coefficients $V_{m,ijkl}$ with $m=1$ to 4. The non-bonded interactions include a Lennard-Jones and a Coulombic term, which are fully included for all those atoms separated by more than 4 sites along the molecular backbone and scaled by a factor 0.5 for the 1-4 pairs. All the Coulombic interactions are scaled by a factor $\alpha_q=0.8$. The Lennard-Jones parameters for the interaction between atoms of different type is calculated using the combination rule $\sigma_{ij}=\sqrt{\sigma_i \sigma_j}$ and $\epsilon_{ij}=\sqrt{\epsilon_i \epsilon_j}$.
In Table \ref{ff} we summarize all the parameters used for the simulations and in Figure \ref{dimerq} the detailed structure of the molecular dimer, as well as the labels for the different atom types and their corresponding charges are displayed. 

\noindent
The Gromacs topology and configuration files are available upon request.

\begin{table}[!h]
\caption{Lennard-Jones, bond, angular and torsional parameters for [PF$_6$]$^-$ and [P$_{1,2,2,4}$] $^+$.}\label{ff1}

\vspace{1cm}
\begin{tabular*}{4.8cm}{p{1.2cm}|p{1.4cm}p{1.8cm}}
   Atom         &   $\sigma$  (\AA)  & $\epsilon$ (kJ/mol)     \\ \hline 
     P              &       3.74        &     0.8368              \\   
     F              &       3.12        &     0.2552              \\ 
     C             &       3.50        &     0.287614         \\
     H             &       2.50        &     0.12552            \\ \hline
\end{tabular*} \hspace{2cm}
\begin{tabular*}{6.3cm}{p{1.2cm}|p{1.7cm}p{4cm}}
     Bond	&	$r_{0,ij}$ (\AA)   & $k_{r,ij}$ (kJ/mol/\AA$^2$)   \\ \hline 
          P   -- F	&	1.606 & 3100     \\  
          P  -- C  &  1.81  & 3550 \\
          C  -- C & 1.529  & 2242 \\
          C  -- H  & 1.09  & 5555.55 \\ \hline
\end{tabular*}

\vspace{1cm}
\begin{tabular*}{7.1cm}{p{1.9cm}|p{2.2cm}p{2.4cm}}
     Angle	&	$\theta_{0,ijk}$ (deg)   & $k_{r,ij}$ (kJ/mol) \\ \hline
P -- F -- P	&	90.0   & 1165     \\
C -- P -- C & 109.5 & 607.8 \\
C -- C -- P   & 115.2 & 509.1 \\
C -- C -- C  &  112.7 & 488.3  \\
H -- P -- C & 110.1 & 389.9 \\
H -- C -- H & 107.8 & 276.1 \\
\end{tabular*}

\vspace{1cm}
\begin{tabular*}{10.4cm}{p{3cm}|p{1.5cm}p{1.5cm}p{1.5cm}p{2.5cm}}
   Dihedral                 &   $V_1$  &  $V_2$  &   $V_3$ & $V_4$  (kJ/mol)  \\ \hline 
     C -- C -- C -- H     &    0.00    & 0.00 & 1.5313 & 0.00 \\
     C -- P -- C -- H     &    0.00    & 0.00 & 0.9270 & 0.00 \\
     C -- P -- C -- C     &    0.00    & 0.00 & 1.1330 & 0.00 \\
     P -- C -- C -- H     &    0.00    & 0.00 & 0.4650 & 0.00 \\
     P -- C -- C -- C     &   -3.2480    & 0.9880 & -0.7150 & 0.00 \\
\end{tabular*}
\label{ff}
\end{table}

\clearpage

\section*{Rotational dynamics}

\noindent
Information of the dynamics of the system has been obtained using rotational self correlation functions, $C(t)=\langle \vec{\delta}_i(t_0).\vec{\delta}_i(t_0+t) \rangle$, of different characteristic vectors defined by two or three atoms depending on the case. When the vectors is defined along a molecular bond, like P3-CP1, two atoms are used to define the direction of the target vector. In other cases, the vector is defined as the normal to the plane defined by three atoms, like in the case of the CH$_3$ groups. 

\noindent
All the $C(t)$ curves were analyzed in the same way: the {\em raw} data was fitted using a stretched exponential of the form:
\begin{equation}
C(t)=C_0 e^{-(t/\tau)^\beta}+\alpha.
\end{equation}
Here, $C_0$ takes care of the sudden drop in the data due to the molecular librations, which decay in a few ps. The additive constant $\alpha$ accounts for the asymptotic value of $C(t)$ that is different than zero in some cases where the rotations are limited to a region of the unit sphere. The relaxation time is then calculated as:
\begin{equation}
\tau^\ast=\int_0^\infty \frac{C(t)-\alpha}{C_0} dt = \frac{\tau}{\beta} \Gamma \left (\frac{1}{\beta} \right ).
\end{equation}

\vspace{2cm}

\noindent{\bf{Internal Relaxation of P$_{1,2,2,4}$}}

\begin{figure*}[!t]
\includegraphics[width=0.45\textwidth]{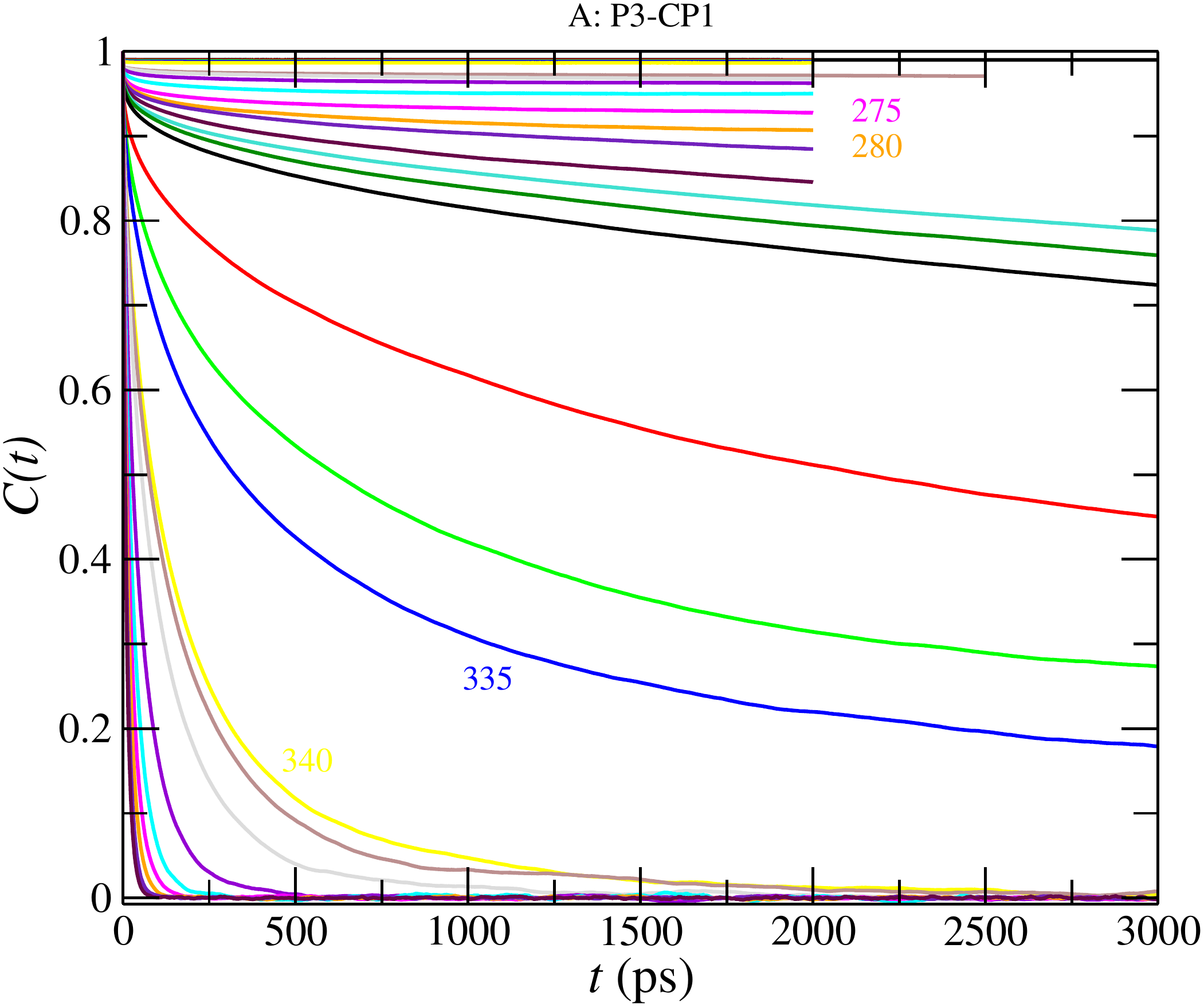}
\includegraphics[width=0.45\textwidth]{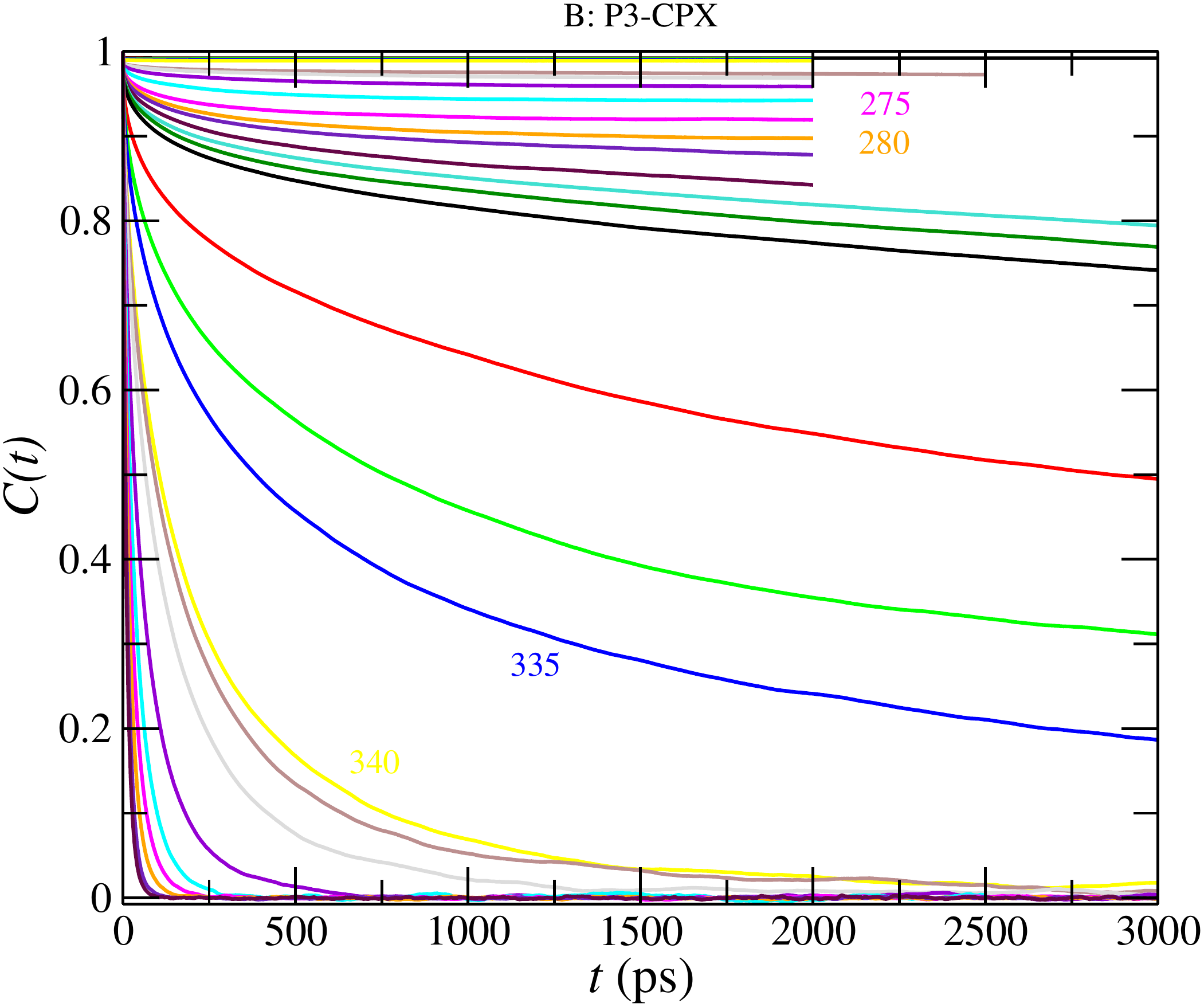}\\ \vspace{1cm}
\includegraphics[width=0.45\textwidth]{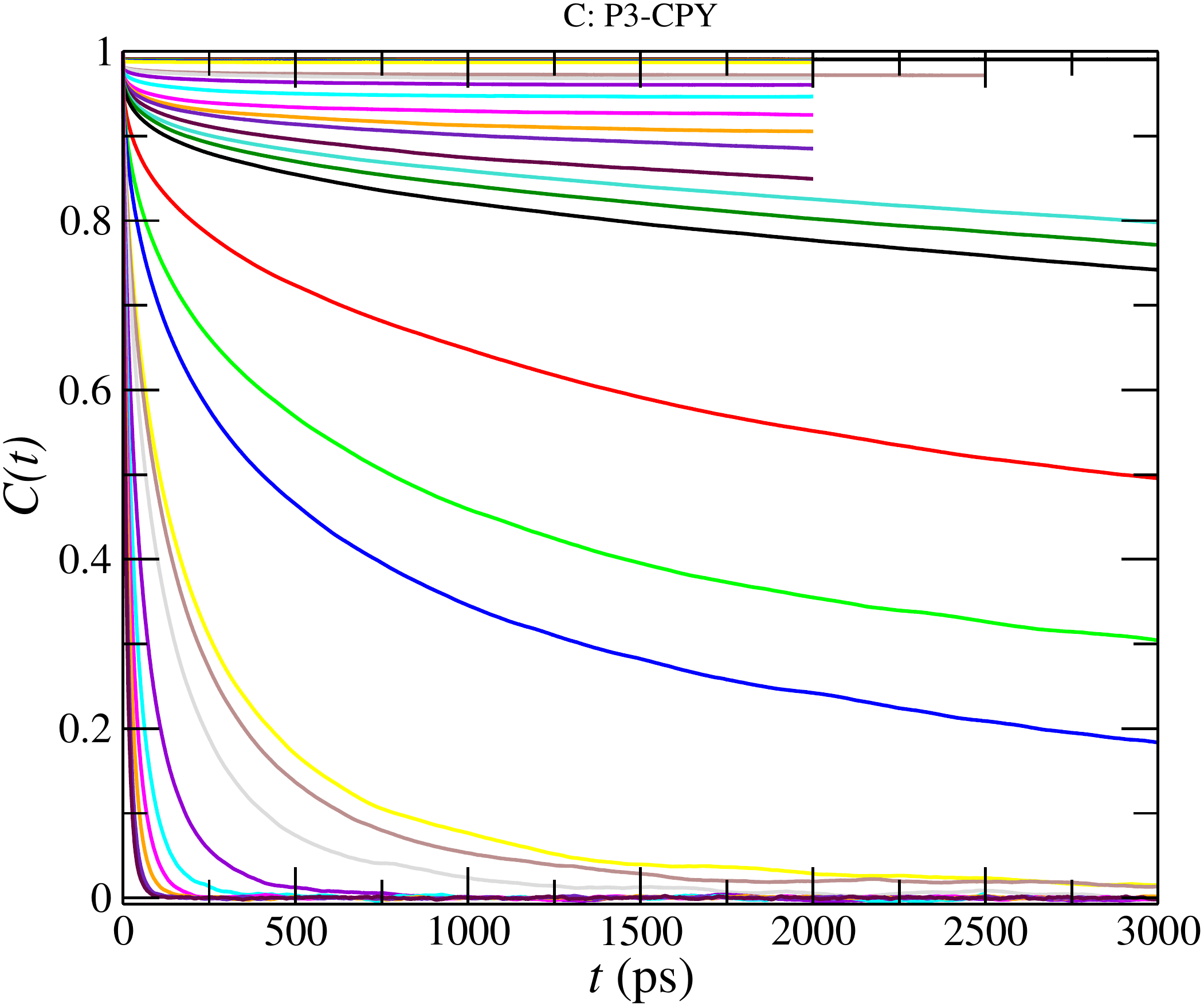}
\includegraphics[width=0.49\textwidth]{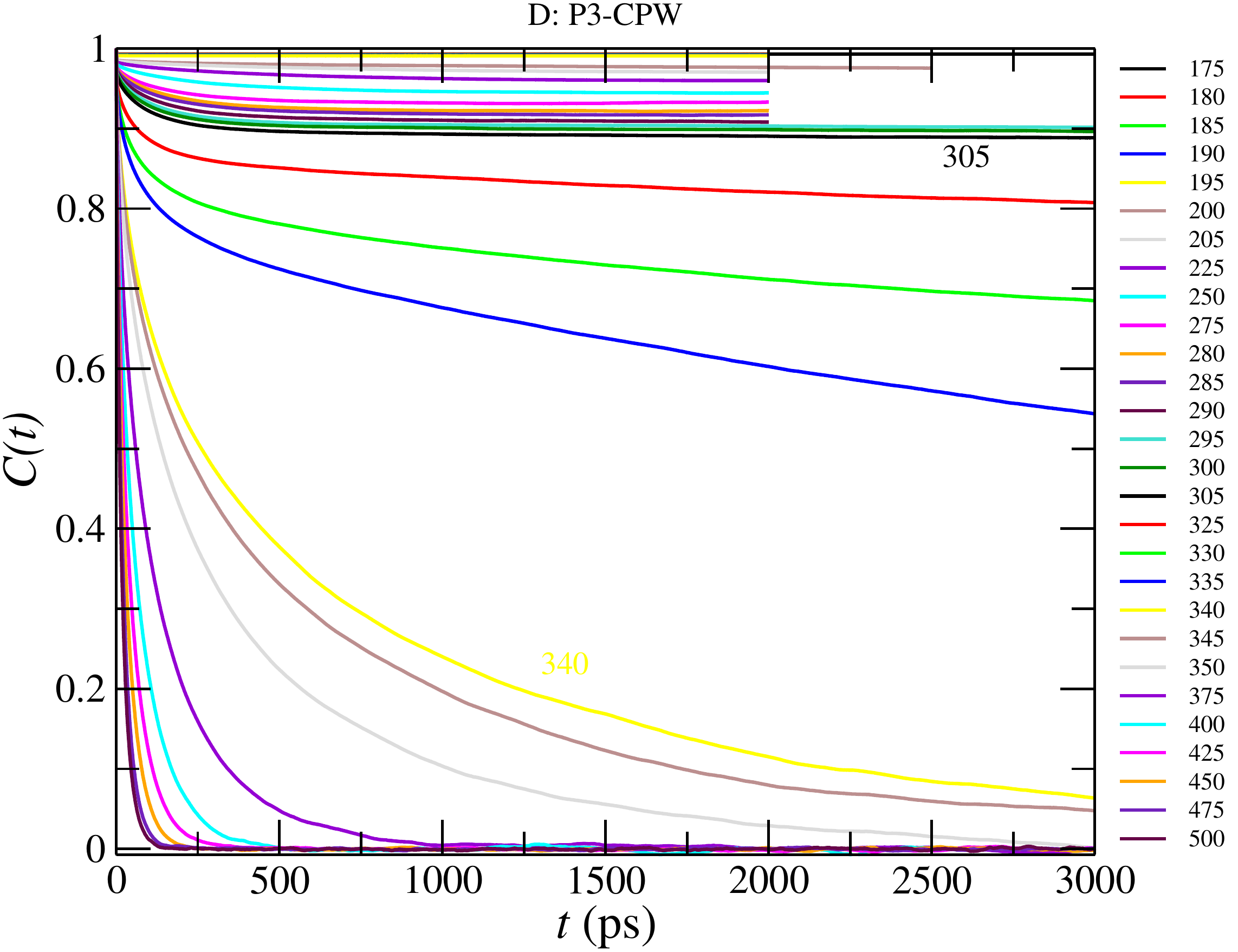}
\caption{Rotational self correlation function involving the innermost atoms of the cation.}
\label{cuatro}
\end{figure*}

\noindent
The rotational self correlation functions involving the central P atom of the cation are shown in Figure \ref{cuatro}. The lower temperatures ($T \le 275$ K) relax to a constant asymptotic value and the higher temperatures ($T \ge 340$ K) relax to zero.
For the cases not involving P3-CPW, the intermediate temperatures display a slow relaxation mode. However, for P3-CPW, the fast relaxation to a constant value extend up to $T \le 305$ K, indicating that for lower temperatures the slow rotations are nearly parallel to the P3-CPW bond.

\begin{figure*}[!t]
\includegraphics[width=0.55\textwidth]{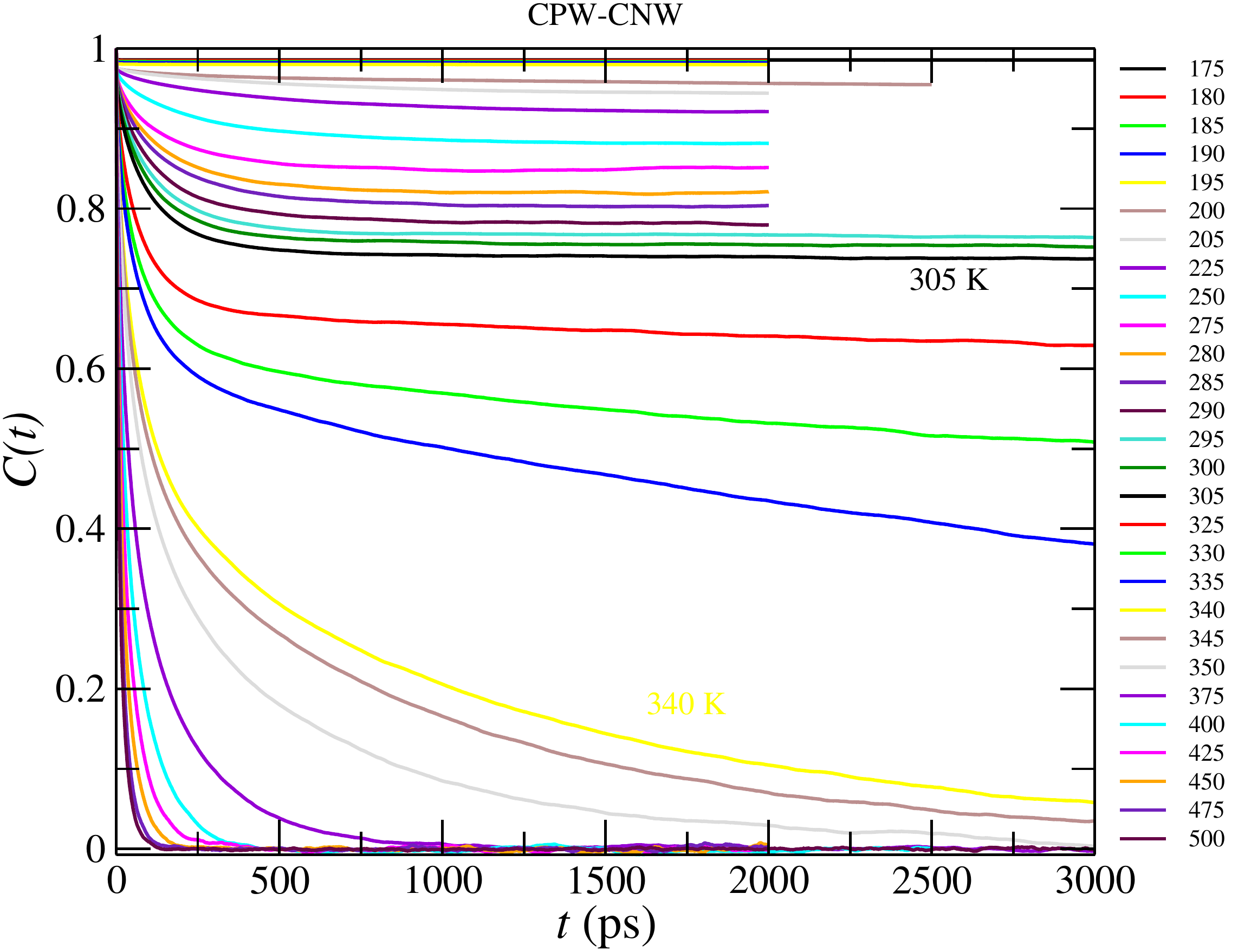}
\caption{Rotational self correlation function for COW-CNW at different temperatures.}
\label{cqq}
\end{figure*}

\clearpage
\noindent
The rotational self correlation function of the vector defined by the CPW-CNW, shown in Figure \ref{cqq} behaves in a similar way as the corresponding curves for P3-CPW, supporting the picture of longitudinal rotations for $T \le 305$ K. 

\begin{figure*}[!t]
\includegraphics[width=0.9\textwidth]{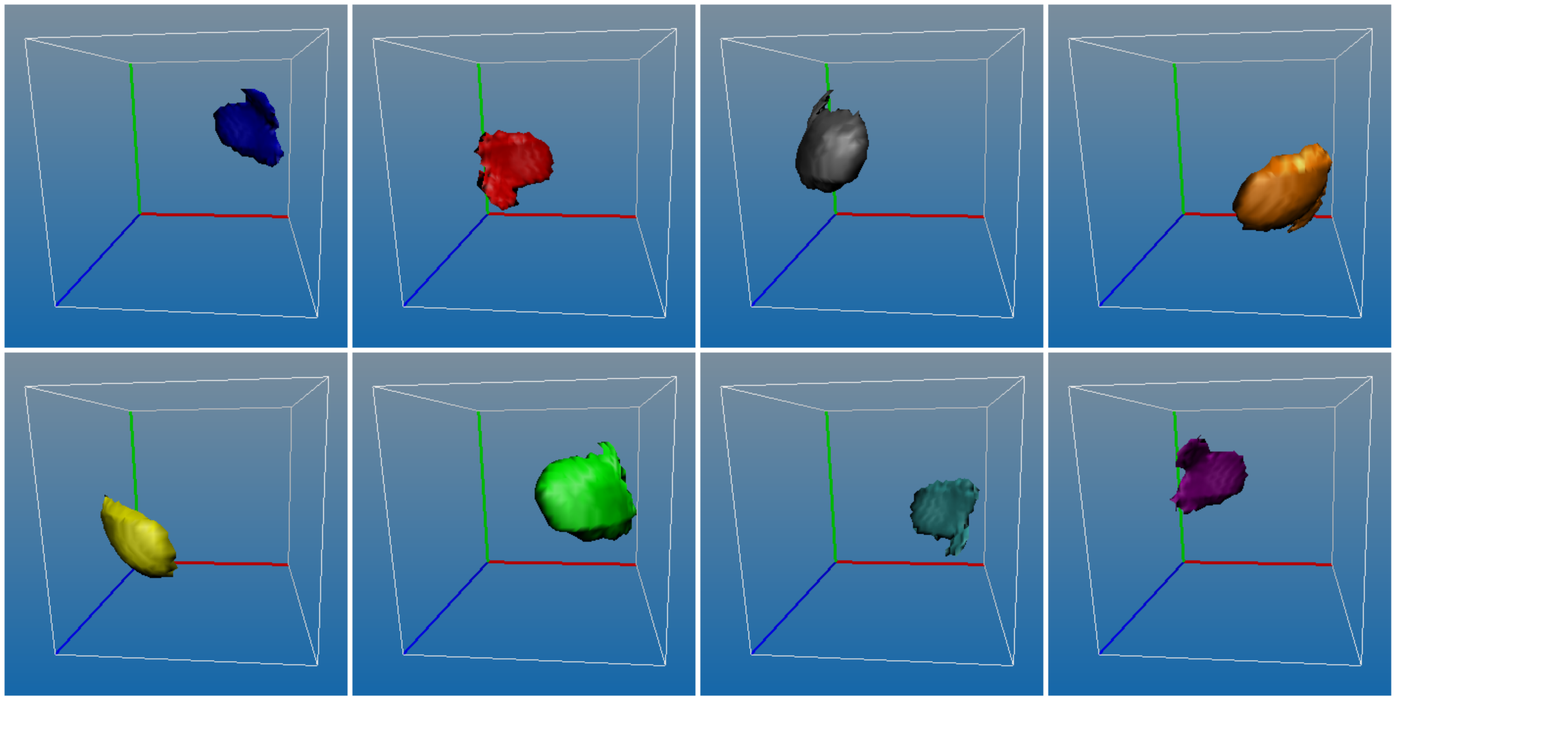} 
\caption{Three dimensional density of the CP1 atom, relative to the position of the corresponding P3 atom at $T=225$ K. The eight different snapshots correspond to the eight non-equivalent molecules in the unit cell.}
\label{cub1}
\end{figure*}

\vspace{2cm}
\noindent
The overall molecular kinetics can be understood by looking at the local density of the CP1 atoms relative to the central P3 atom. Since the unit cell has eight equivalent molecules with different orientations, we average the densities of the corresponding molecules across different unit cells. The simulation box contains 64 unit cells and in each unit cell there are 8 ion pairs, totaling 512 ion pairs. Then we can use a labeling scheme with two indexes: $c_i, m_j$, with $i=1, \cdots, 64$ and $j=1, \cdots, 8$. More explicitly, $c_i$ indicates the unit cell to which a molecule corresponds, and $m_i$ is the molecule label in the unit cell. In this way we calculate eight different density maps, one for each molecule $m_j$ in the unit cell, by averaging across the 64 different unit cells. In Figure \ref{cub1} we show the eight different densities of CP1 relative to P3 calculated with this method.

\end{document}